\documentclass[10pt,twocolumn,letterpaper]{article}

\usepackage{wacv}      

\usepackage{booktabs} 
\usepackage{gensymb}
\usepackage{amsmath}
\usepackage{amsfonts}
\usepackage{enumitem}
\usepackage[ruled]{algorithm2e}
\usepackage{soul}
\usepackage{microtype}
\usepackage{units}
\usepackage{gensymb}
\usepackage{array}
\usepackage{subcaption}
\usepackage{xcolor}
\usepackage{pifont}
\usepackage{multirow}
\usepackage{float}
\usepackage[dvipsnames]{xcolor}
\usepackage{bm}
\usepackage{rotating}
\usepackage{xspace}
\definecolor{ForestGreen}{RGB}{34,139,34}
\usepackage{diagbox}

\newcommand{\figref}[1]{Figure\ \ref{#1}}
\newcommand{\secref}[1]{Section\ \ref{#1}}

\newcommand{\tabref}[1]{Table\ \ref{#1}}

\definecolor{wacvblue}{rgb}{0.21,0.49,0.74}
\usepackage[pagebackref,breaklinks,colorlinks,allcolors=wacvblue]{hyperref}
\usepackage{doi}

\def\wacvPaperID{2001} 
\def\confName{WACV}
\def\confYear{2026}

\title{DreamAnywhere: Object-Centric Panoramic 3D Scene Generation}

\author{
    Edoardo A. Dominici\textsuperscript{1} \quad
    Jozef Hladký\textsuperscript{1} \quad
    Floor Verhoeven\textsuperscript{1} \quad
    Lukas Radl\textsuperscript{2} \\
    Thomas Deixelberger\textsuperscript{1}\textsuperscript{2} \quad
    Stefan Ainetter\textsuperscript{1} \quad
    Philipp Drescher\textsuperscript{1}\textsuperscript{2} \quad
    Stefan Hauswiesner\textsuperscript{1} \\
    Arno Coomans\textsuperscript{1} \quad
    Giacomo Nazzaro\textsuperscript{1} \quad
    Konstantinos Vardis\textsuperscript{1} \quad
    Markus Steinberger\textsuperscript{1} \textsuperscript{2} \\ \\
    \textsuperscript{1}Huawei Technologies \qquad
    \textsuperscript{2}Graz University of Technology
}

\begin{document}

%

\twocolumn[{%
\renewcommand\twocolumn[1][]{#1}%
\maketitle
\includegraphics[width=\textwidth]{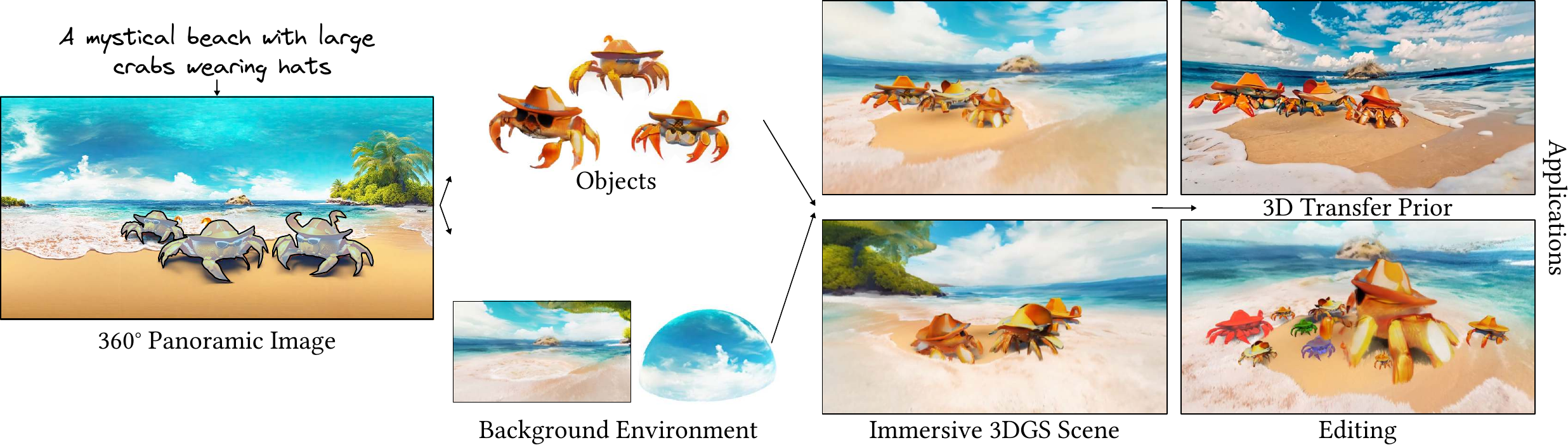}
\captionof{figure}{Starting from a text prompt, our method generates 3D scenes leveraging a 360\degree{} panoramic image as intermediate representation to extract, reconstruct and compose objects in the environment. Our scenes,  represented as 3D Gaussian splats, support long-range exploration and have high structural coherence even under large camera offsets. This improves immersive scene generation for existing applications, enables intuitive object-level editing and makes them strong 3D priors for world-to-world transfer models \cite{cosmos}. \vspace{1em}}
\label{fig:teaser}
}]

Recent advances in text-to-3D scene generation have demonstrated significant potential to transform content creation across multiple industries. Although the research community has made impressive progress in addressing the challenges of this complex task, existing methods often generate environments that are only front-facing, lack visual fidelity, exhibit limited scene understanding, and are typically fine-tuned for either indoor or outdoor settings. 
In this work, we address these issues and propose \textit{DreamAnywhere}, a modular system for the fast generation and prototyping of 3D scenes. 
Our system synthesizes a 360° panoramic image from text, decomposes it into background and objects, constructs a complete 3D representation through hybrid inpainting, and lifts object masks to detailed 3D objects that are placed in the virtual environment. 
\textit{DreamAnywhere} supports immersive navigation and intuitive object-level editing, making it ideal for scene exploration, visual mock-ups, and rapid prototyping---all with minimal manual modeling. These features make our system particularly suitable for low-budget movie production, enabling quick iteration on scene layout and visual tone without the overhead of traditional 3D workflows.
Our modular pipeline is highly customizable as it allows components to be replaced independently. 
Compared to current state-of-the-art text-\ and image-based 3D scene generation approaches, \textit{DreamAnywhere} shows significant improvements in coherence in novel view synthesis and achieves competitive image quality, demonstrating its effectiveness across diverse and challenging scenarios.
A comprehensive user study demonstrates a clear preference for our method over existing approaches, validating both its technical robustness and practical usefulness.

\section{Introduction}
Generating large and immersive virtual 3D scenes from text remains a significant challenge with transformative applications in movie production, gaming, virtual reality, and simulation.
Recent advances in generative modeling have led to impressive results but the 3D scenes are only coherent within small ranges of camera movement, resulting in incomplete structures and hallucinations when viewers attempt to explore beyond narrow boundaries.

Existing text-to-3D approaches leverage the richness and stylistic diversity of internet-scale diffusion models for progressive outpainting \cite{hoellein2023text2room} or distillation sampling \cite{shriram2024realmdreamertextdriven3dscene, dreamsceneLi2024}, but struggle to capture global coherence, leading to increasing semantic drift and structural collapse due to the difficulty of integrating depth estimates into a single 3D scene representation. Video diffusion models have been combined with large reconstruction models \cite{yu2024viewcraftertamingvideodiffusion, chen2025flexworldprogressivelyexpanding3d} but still face limitations in generating complete scenes observable from vastly different viewpoints maintaining coherency.
Panoramic 360\degree{} images are an inherently good starting point for immersive viewing, in fact \cite{li2024scenedreamer360, zhou2024holodreamerholistic3dpanoramic} combine panoramic images with inpainting to enable parallax effects when moving away from the origin, further extended to multiple layers by \cite{yang2024layerpano3d}.  
Similarly, approaches designed for 2D fly-throughs \cite{infinite_nature_2020,scenescape2023,yu2023wonderjourney,yu2024wonderworld} can create visually convincing animations, but lack the global 3D structure required for full scene exploration. 

Whilst large-scale and artistic 3D scene data remains scarce, 
there exists a wealth of 3D data for individual objects, offering a valuable resource for compositional approaches. Moreover, decomposing a 3D environment into individual entities offers important advantages for downstream applications~\cite{Factoring2018,armeni20193d}. If the layout and description of objects in a scene are known or provided manually, immersive scenes can be generated~\cite{dreamsceneLi2024}. However, optimizing scene layouts end-to-end continues to be a major challenge \cite{zhang2023scenewiz3d,epstein2024disentangled}.

Building on these insights, we present \textit{DreamAnywhere}, a text-to-3D generation framework that leverages 360\degree{} panoramic image generation to drive scene style and geometry, while decomposing scenes into individual objects to preserve structural coherence and enable the generation of navigable 3D environments. We make the following contributions:
\begin{itemize}[leftmargin=*]
\item A novel unified and modular system for generating 3D scenes with high 3D coherence and visual fidelity while facilitating interactivity.
\item A panoramic image generation method that jointly fine-tunes a 360\degree{} image diffusion model with a perspective conditioning mechanism, improving out-of-domain sampling for creative applications.
\item A high-quality object reconstruction method that first synthesizes a detailed reference image from multimodal cues---textual, geometric, and stylistic---and then leverages it to guide robust 3D model generation and alignment, even for low-quality and incomplete segmented inputs.
\item A hybrid inpainting strategy that combines 2D and 3D techniques: large-scale holes resulting from object removal are inpainted in the 360\degree{} image for global coherence, while smaller disocclusions caused by the 3D projection are addressed with 3D inpainting.
\end{itemize}




To evaluate our approach, we conduct both quantitative comparisons and a controlled user study, demonstrating that our method achieves superior quality, especially under large view offsets.
Furthermore, qualitative results and user feedback indicate that our approach yields much more coherent and immersive scenes. As shown in \figref{fig:teaser}, this holds even for diverse camera movements.

\begin{figure*}[t]
    \centering
    \includegraphics[width=\textwidth]{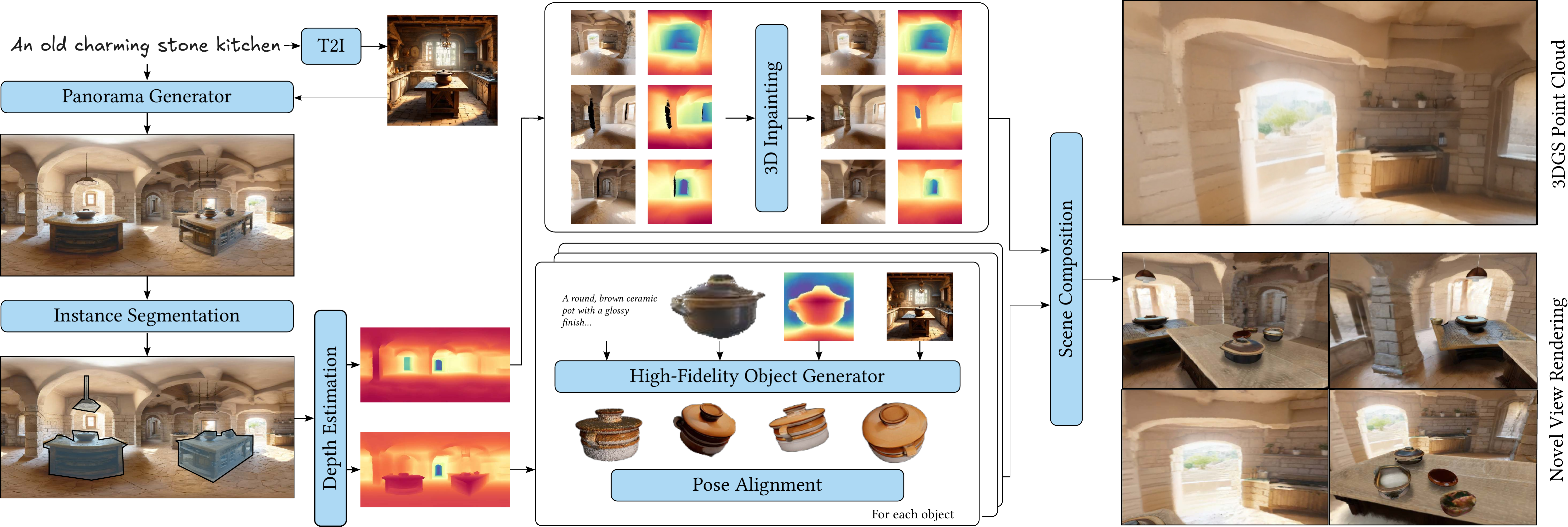}
     \caption{An overview of our architecture. We start by generating a 360\degree~panorama, followed by instance segmentation to separate the foreground objects from the scene background. Segmented objects are lifted to 3D using semantic and geometric priors, while a 3D inpainting process converts the image to a 3DGS representation and fills disoccluded regions. Finally, the scene is composited, enabling navigation of the 3D environment at interactive rates.} 
    \label{fig:overview}
\end{figure*}

\section{Related Work}
Our 3D scene generation work lies at the intersection of image-based generation and compositional scene synthesis, bridging recent advances in both areas. For an in-depth discussion we refer readers to recent surveys on scene generation ~\cite{surveygen2023, surveyscene20254} and object generation ~\cite{surveygen2023, po2024state}. 

\subsection{Image-based Scene Generation}

A successful line of work in image-based scene generation focuses on monolithic representations, where the entire scene is captured in a single unified 3D structure. 
A common technique involves iterative outpainting with image diffusion models to progressively extend the scene along a camera trajectory \cite{scenescape2023,hoellein2023text2room,yu2023wonderjourney,yu2024wonderworld,Liang_2024_CVPR}. 
Although effective to some extent, these methods often suffer from the accumulation of visual artifacts and distortions across iterations, and offer limited flexibility for isolating and manipulating individual scene components.

An alternative set of techniques reconstructs 3D scenes from 2D images using inpainting-based methods, which lift partial views into 3D and fill occluded regions through inpainting~\cite{shriram2024realmdreamertextdriven3dscene,Zhang2023Text2NeRFT3}. 
Although these reconstructions can appear photorealistic from specific angles, their quality tends to degrade when viewed from arbitrary viewpoints.

More recently, panorama-based approaches such as \mbox{PanoDreamer} \cite{paliwal2024panodreamer} and PanoDiffusion \cite{wu2023panodiffusion} attempt to synthesize entire 360\degree{} panoramic images with depth information through in- or outpainting. 
Rather than iteratively expanding standard 2D images, other methods focus on creating 360\degree{} panorama images in a single step~\cite{MVDiffusion2024,li2024scenedreamer360,zhou2024holodreamerholistic3dpanoramic,stan2023ldm3dvr}.
Another area of research addresses the handling of occlusions via multi-view generation~\cite{ye2024diffpano}, layered approaches~\cite{yang2024layerpano3d} or explicitly defining the corresponding entities~\cite{zhou2025dreamscene360}. 
Video diffusion models have strong temporal coherency but are still limited in generating 3D content that is consistent along long camera trajectories ~\cite{yu2024viewcraftertamingvideodiffusion,ren2025gen3c3dinformedworldconsistentvideo,chen2025flexworldprogressivelyexpanding3d,zhai2025stargenspatiotemporalautoregressionframework,SceneSplatter2025, voyager}. \
While image-based methods offer strong visual quality and global coherence, they generally lack fine-grained control over individual scene elements and are limited in navigational flexibility, especially when compared to compositional scene generation approaches, which we discuss in the following section.

\subsection{Compositional Scene Generation}
In contrast to monolithic approaches, compositional scene generation focuses on constructing scenes from multiple, independently generated objects which can be edited or animated. 
Recent advances in this area predominantly leverage pre-trained image diffusion models for scene generation \cite{huang2024midi,Dogaru2024Gen3DSR,zhou2024DeepPriorAssembly,han2024reparocompositional3dassets,chen2025comboverse}. The key differences between these methods lie in the strategies used to layout the scene.

Image-guided methods~\cite{huang2024midi,Dogaru2024Gen3DSR,zhang2023scenewiz3d,zhou2024DeepPriorAssembly,han2024reparocompositional3dassets} use generated or real images to infer each object's location and orientation in the scene. Similarly, distillation-based approaches \cite{epstein2024disentangled,dreamsceneLi2024,chen2025comboverse} employ diffusion models to guide object placement. However, since these methods typically rely on single (non-panoramic) images, they are often limited in terms of scale and global coherence of the generated scenes.

To overcome these limitations, proxy-guided techniques introduce explicit control over scene structure, for example by letting users define the scene structure through various semantic annotations or by placing proxy geometries \cite{blockfusion,schult24controlroom3d,fang2023ctrl,Cohen-Bar_2023_ICCV,yang2024scenecraft}. While these techniques provide great flexibility in terms of scale and coherence, they typically require extensive user input and manual scene design.

Overall, compositional approaches provide a powerful alternative to image-based scene generation, trading off the ease of automation for improved control, reusability, and interpretability of individual scene elements.

\section{Method}
The DreamAnywhere architecture proceeds through three stages as illustrated in \figref{fig:overview}. In the \textbf{first stage}, a $360\degree{}$ panoramic image $I_F$ is generated from a text prompt and stylized by an input perspective image. From $I_F$, foreground objects are identified, segmented, and erased, producing a \textit{clean} background image $I_B$. Then, the depth of $I_B$ is aligned to that of $I_F$, producing a single, unified depth map and laying the groundwork for subsequent 3D composition. The \textbf{second stage} focuses on generating high-fidelity 3D Gaussian Splatting (3DGS) models of the segmented objects and computing their transformations for accurate placement in the 3D scene. In parallel, the \textbf{third stage} converts the background image $I_B$ into a high-resolution 3DGS point cloud. During this process, disoccluded regions are inpainted to ensure visual consistency across viewpoints under novel camera offsets. Finally, the outputs of the second (objects) and third (background) stages are fused into a unified 3DGS scene, enabling immersive and interactive navigation.
The remainder of this section presents all the components of our system, emphasizing the most essential technical and implementation details. Additional information and extended ablation studies regarding 2D and 3D inpainting, object and background generation, pose estimation and scene compositing are available in the supplementary material. 

\subsection{Image Generation}
\label{subsec:panorama_generation}

\subsubsection{360\degree{} Panorama Diffusion}
\begin{figure}[t]
    \centering
\includegraphics{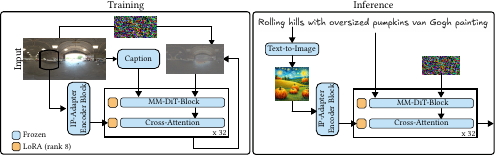}
    \caption{Our image generation pipeline uses a diffusion model to denoise an equirectangular image \cite{yang2024layerpano3d, feng2023diffusion360}, but additionally combines perspective image conditioning through a decoupled cross-attention layer~\cite{ye2023ip}. We train an equirectangular LoRA jointly with the perspective pre-trained cross attention layer, by encoding random perspective crops of the panorama, showing stronger generalization capabilities to out-of-domain panorama sampling.}
    \label{fig:panogen}
\end{figure}

The initial stage of our pipeline generates the 360\degree{} panorama image $I_{F}$ from a user-provided text prompt. Existing text to panorama diffusion models are typically LoRA-fine-tuned on small, less-diverse equirectangular datasets, limiting their ability to generalize to out-of-distribution styles/compositions despite these capabilities existing in the pre-trained base model. 
To improve generalization and style fidelity, we guide the generation process using a perspective image derived from the same prompt, providing soft conditioning without enforcing pixel-level alignment. We achieve this by using an IP-Adapter-style mechanism \cite{ye2023ip} that introduces separate cross-attention layers in all transformer blocks of the diffusion model. We observe that simply adding this mechanism post-hoc to a panoramic LoRA model is ineffective due to a mismatch in perspective (for IP-Adapter) and panoramic (for LoRA) training distributions as shown in the ablation in \figref{fig:ablation}. Instead, we \textit{jointly} fine-tune the panoramic LoRA with the IP-adapter using random perspective renders of the equirectangular image (\figref{fig:panogen}), enabling effective style transfer from perspective to panoramic images.

We base our model on Stable-Diffusion-3.5-Large and fine-tune a rank-8 LoRA on top of all attention blocks on a mix of indoor \cite{Matterport3D} and outdoor \cite{polyhaven} images. We train the model for 10k iterations with a learning rate of $10^{-5}$ and a batch size of 4.

\subsubsection{Instance Segmentation}
\label{subsec:instance_seg}
To identify suitable candidates for object reconstruction in the $360\degree{}$ images, we employ the off-the-shelf open-set segmentation pipeline \textit{Grounded-SAM} \cite{ren2024groundedsamassemblingopenworld}. To improve robustness in stylized or out-of-distribution scenes, we replace the original class identification stage (RAM~\cite{zhang2023ram}) with a larger vision-language model (GPT-4V), improving category recognition under diverse visual conditions.
Given the limitations of current object generators, we exclude categories that are not good candidates for object reconstruction and are typically considered background elements (such as grass, walls, and floors). To further ensure high-quality segmentation, we sort all detected objects by their predicted IoU confidence and retain only the highest-scoring masks.
For each retained object, the pipeline outputs a class identifier, a textual description, and a fine-grained segmentation mask, which are later used in \secref{subsec:object_generation}.


\subsubsection{2D Inpainting}
We generate the background panorama $I_{B}$ by erasing the foreground objects identified in the full panorama using the union of their segmentation masks. To synthesize plausible new content in the masked regions, we employ a diffusion-based inpainting model, Dreamshaper-8-Inpainting \cite{dreamshaper2023}, chosen for its ability to synthesize meaningful background content. We reduce hallucinations and artifacts in a two-stage process. First, we pre-inpaint the masked regions using LaMA~\cite{Lama2022} to obtain a coherent object-free initialization. Second we feed this through a custom LoRA fine-tuned on panoramic images from the Structured3D dataset \cite{Structured3D} for indoor defurnishing (in a similar spirit to \cite{matterport2024defurnishing}).
To further improve image quality, we apply standard techniques: the segmentation masks are slightly dilated, wrap-around consistency is enforced via horizontal padding (256 pixels), and Poisson blending seamlessly integrates the original panorama with the newly inpainted content.
Although we experimented with repurposing the panoramic LoRA for inpainting, we found that a dedicated, specialized pipeline significantly reduces hallucinations without requiring manual intervention, especially in indoor scenes.
The resulting empty panorama image is then lifted to 3D to form the background point cloud as described in \secref{subsec:background_generation}.

\subsubsection{Depth Estimation}
We employ different depth estimation strategies for indoor and outdoor scenes due to their distinct geometric challenges. For indoor scenes, we use EGformer \cite{EGformer2023}, which excels in predicting the axis-aligned structure typical of indoor environments. For outdoor scenes, we instead use the \mbox{360MonoDepth} framework \cite{depreyarea2021360monodepth}, which aligns multiple perspective depth estimates obtained with Depth-Anything-V2 and possesses strong zero-shot generalization capabilities. To address unreliable depth predictions in low-detail regions, such as the sky, we explicitly detect and segment these areas using GroundedSAM \cite{ren2024groundedsamassemblingopenworld} and assign the maximum depth value.
By construction, we obtain two independent depth estimates for $I_F$ and $I_B$ and perform an alignment step from $I_B$ to $I_F$ to ensure that objects are placed properly in their corresponding locations. Similarly to concurrent work, we find that a global affine transform is not sufficient and instead employ a per-pixel shift and global scale that minimizes the $L2$ depth error in overlapping regions between the images, subject to as-rigid-as-possible constraints \cite{ARAP}.

\subsection{High Quality Object Generation}
\label{subsec:object_generation}
\begin{figure}
    \centering
    \includegraphics{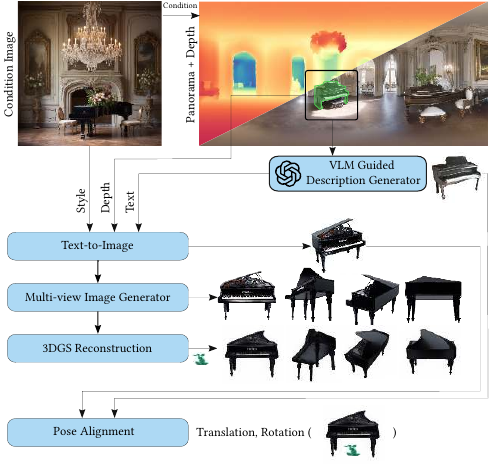}
    \caption{Our object reconstruction pipeline leverages the panorama and style information to generate a high-resolution reference image to be used for multi-view generation. The generated multi-view images are then transformed into 3D Gaussian splats through a reconstruction pipeline. Finally, we align the generated object with the original and place it in the scene.} 
    \label{fig:overview_objectgen}
\end{figure}

To enable stylistically aligned and geometrically consistent 3D object reconstructions, we condition image-to-3D models using RGBD data derived from the segmented panorama.
However, segmented object images are often incomplete, low-resolution, seen from suboptimal viewpoints, or contain generative artifacts (such as distortions or incorrect limbs).
To overcome these problems, we introduce a resynthesis step that produces a high-quality reference image of the object, utilizing both semantic and geometric cues. This image is then used to guide the 3D reconstruction of each object.
Our pipeline consists of three stages: reference image generation, 3D object generation, and pose estimation for accurate placement in the scene. We illustrate this process in \figref{fig:overview_objectgen}.

\subsubsection{Reference Image Generation}
\label{sec:referimg_objgen}
\begin{figure}
    \centering
    \includegraphics{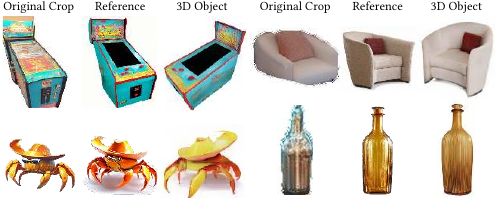}
    \caption{Qualitative evaluation of our object reconstruction module. Employing multimodal cues faithfully captures the stylistic and structural essence of low-fidelity or incomplete objects, reconstructing a suitable 3D counterpart to be later placed in the environment.} 
    \label{fig:ablation_object}
\end{figure}

We synthesize a high-resolution reference image for each object by conditioning a diffusion model on the available multimodal cues.
First, we extract a textual description of the object by prompting a vision-language model with a perspective view of a crop containing the object. This includes a description of the object's visual attributes such as pose and color, serving as the primary conditioning signal. 
Second, we compute a partial depth map by combining the object's segmentation mask with the estimated depth, offering geometric guidance for shape.
Finally, we include a style image, reused from the panorama generation process, to ensure visual consistency and thematic coherence across the scene.
As shown in \figref{fig:ablation_object}, our approach results in realistic and style-consistent object views.
This approach is better suited for 3D reconstruction compared to other diffusion-based amodal completion methods (see \figref{fig:ablation_resynthesis}).

\subsubsection{Image-based Object Generation}
\label{sec:imgobjgen}
Our object reconstruction pipeline first generates six multi-view images at fixed camera poses, and then transforms them into a NeRF representation that captures both the geometry and color of the object. We found NeRF to be more robust than 3DGS as a differentiable representation when learning the reconstruction. Therefore, we opt to transform it into 3DGS in a separate post-processing step.
We leverage Zero123++ \cite{shi2023zero123plus} for the multi-view generation task and use the triplane decoder of InstantMesh \cite{xu2024instantmesh} to convert the multi-view images into NeRF. In practice, generative models often introduce artifacts, including transparency, incomplete regions or structural inconsistencies.
To enable a reliable and automated object reconstruction pipeline, we introduce a single-round refinement stage after generation.
During this stage, we generate multiple object candidates and rank them by visual quality and alignment with the original input image using a VLM, similar to~\cite{zhou2025dreamscene360}. The VLM selects the best result from five candidates, one of which is generated using only the original cropped object image.

\subsubsection{Pose Estimation}
\label{subsec:object_alignment}
We decompose object pose estimation into two parts: the absolute pose of the original object $O_\text{crop}$ with respect to the viewing origin and the relative transformation between $O_\text{crop}$ and its high-quality counterpart $O_\text{hq}$. 
The absolute pose is estimated by projecting the crop's midpoint along a ray from the camera origin to the estimated mean depth. Orientation and scale are heuristically derived based on the crop geometry and panorama alignment. This provides a coarse but coherent placement aligned with the image-plane and panorama geometry.
To semantically align the point cloud of the high-quality object $O_\text{hq}$ with the original $O_\text{crop}$ we compute the relative pose between them. We employ MAST3R~\cite{mast3r_arxiv24} on image pairs (perspective projections) of the original and high-quality object to obtain the relative pose.
The final pose of $O_\text{hq}$ preserves semantic alignment with the original object's placement.

\subsection{3D Background Generation}\label{subsec:background_generation}
\begin{figure}[t]
    \centering
\includegraphics[width=\linewidth]{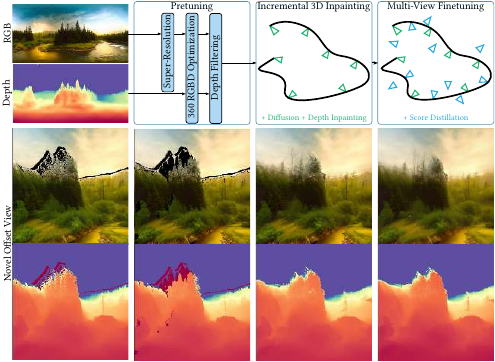}
    \caption{The background generation stage starts by identifying large occluded regions and preparing the 3DGS point cloud for inpainting, then proceeds to incrementally fill in new content and finally enforces multi-view consistency when rendering from novel views with score distillation.
    } 
    \label{fig:3d_inpaint}
    \vspace{-0.1em}
\end{figure}

Generating a 3D scene suitable for exploration from a single 360\degree{} image is an underconstrained problem, as the available information is limited to a single viewpoint. To address this, we employ a three-step process that reconstructs missing scene content through a combination of inpainting and score distillation sampling (SDS) \cite{poole2023dreamfusion}. First, we optimally initialize the 3DGS point cloud from the panoramic image and prepare disocclusions for inpainting through a pretuning stage; Second, we incrementally populate disocclusions with new Gaussians through a 3D inpainting process conditioned on the existing point cloud; Finally, we ensure multi-view consistency by applying SDS on a larger set of possible views. During all optimization steps, we leverage MCMC-densification~\cite{kheradmand2024mcmc}, a process demonstrated in \figref{fig:3d_inpaint}.

\subsubsection{Background Initialization and Pretuning.}
We initialize the 3DGS point cloud by unprojecting the pixels through a spherical camera placed at the origin, where the covariance matrices are initialized from normals derived from the depth map $D_B$, resulting in pixel-tight views when viewed from the origin. 
For pretuning, we optimize the initial Gaussians using mixed supervision from an upscaled version of the background panorama $I_B$ (90\%) and random views (10\%). The $I_B$ loss follows standard 3DGS supervision, while the random-view supervision focuses on identifying disoccluded regions.

\vspace{-0.25em}

\subsubsection{Incremental Inpainting}
To synthesize realistic new content in disoccluded regions, we leverage a pre-trained inpainting model conditioned on the rendered projections of the existing 3DGS splats. Specifically, we select a set of keyframes that target important regions with missing content. For each keyframe we render the point cloud and generate an inpainted RGB image and an inpainting mask, computed by applying max-pooling (kernel size 25) and thresholding ($T{<}0.5$) on the transmittance map $T$, isolating low-opacity regions as targets for inpainting.
For each pixel in the target region, we instantiate a new Gaussian by unprojecting it with a scene-aligned depth estimate. This depth is obtained using a diffusion model repurposed and fine-tuned for 3DGS depth inpainting \cite{liu2024infusion}, and provides a strong geometric prior. The inpainting process is done incrementally per keyframe, minimizing inconsistencies across overlapping views, similarly to \cite{hoellein2023text2room}.

\subsubsection{Multi-view Fine-tuning}
Finally, to consolidate the inpainted views with the existing 3D content, we employ mixed supervision from the panorama image (25\%), the inpainted images (25\%) and SDS~\cite{poole2023dreamfusion} from random views around the inpainted regions (50\%).
This final stage of SDS fine-tuning helps to remove view-inconsistencies and ensures that the Gaussians introduced during inpainting are consistent and aligned with the existing scene as shown in \figref{fig:3d_inpaint}.

\subsection{Composition}
The last stage of the pipeline composes the final scene, by placing the 3D objects onto the background point cloud's world frame, resulting in a navigable 3D scene. Since the scene consists of a set of splat parameters and their explicit transformations, post-processing effects and re-positioning of objects can be applied trivially, a key strength of the proposed system.

To further improve visual coherence, we address two key sources of error that can disrupt immersion. The first is the elimination of contact shadows---an unavoidable consequence of the mask dilation step used during object removal. Since achieving full control over lighting in diffusion models \cite{lighting0, lighting1} and relighting remain challenging problems, we reintroduce shadows through a splat-aware shadow mapping pass (details in the supplemental). The second issue involves inaccurate pose estimation, which can lead to objects violating physical plausibility. To mitigate this, we automatically snap object positions to dominant support planes (e.g. floor or walls), ensuring a more realistic and physically grounded scene integration.

\vspace{-0.1em}

\section{Results}
\label{sec:results}
\begin{figure}[t]
  \centering
  \setlength{\tabcolsep}{1.0pt}
  \renewcommand{\arraystretch}{1.0}
   \begin{tabular}{@{}lccc@{}} 
    \rotatebox{90}{\footnotesize{\hspace{2.1em}Ours}}
      & \includegraphics[width=0.3\linewidth]{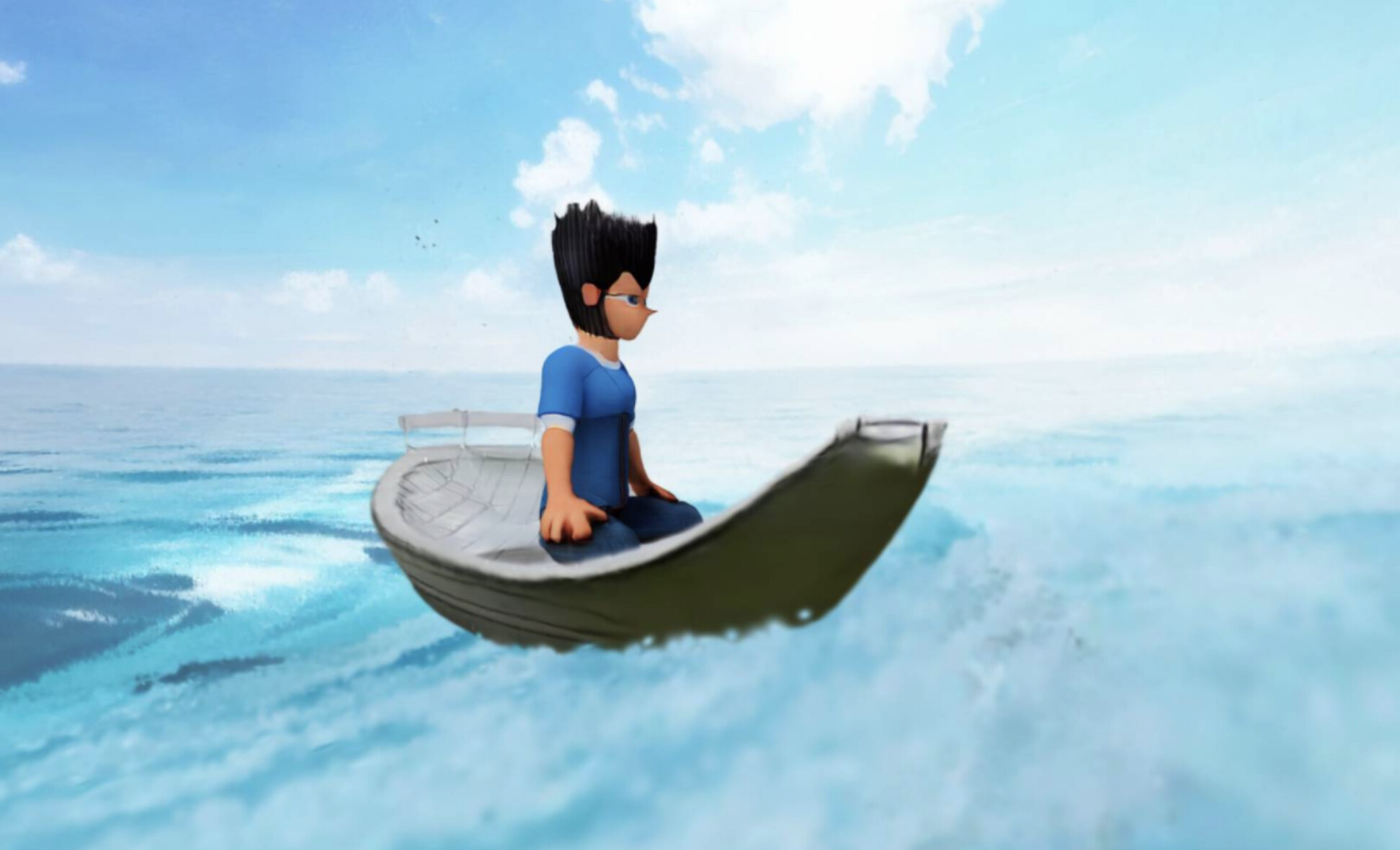}
      & \includegraphics[width=0.3\linewidth]{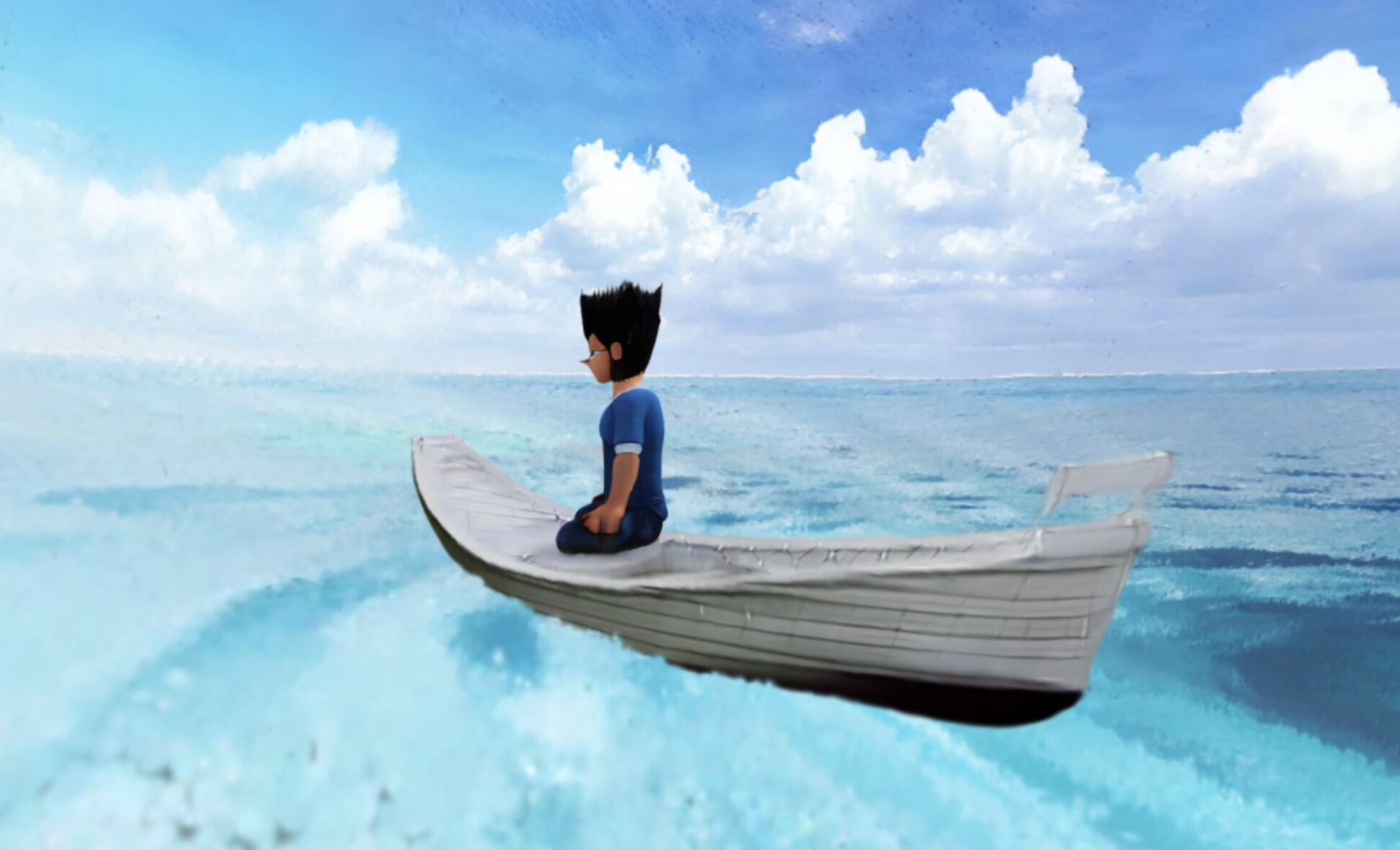}
      & \includegraphics[width=0.3\linewidth]{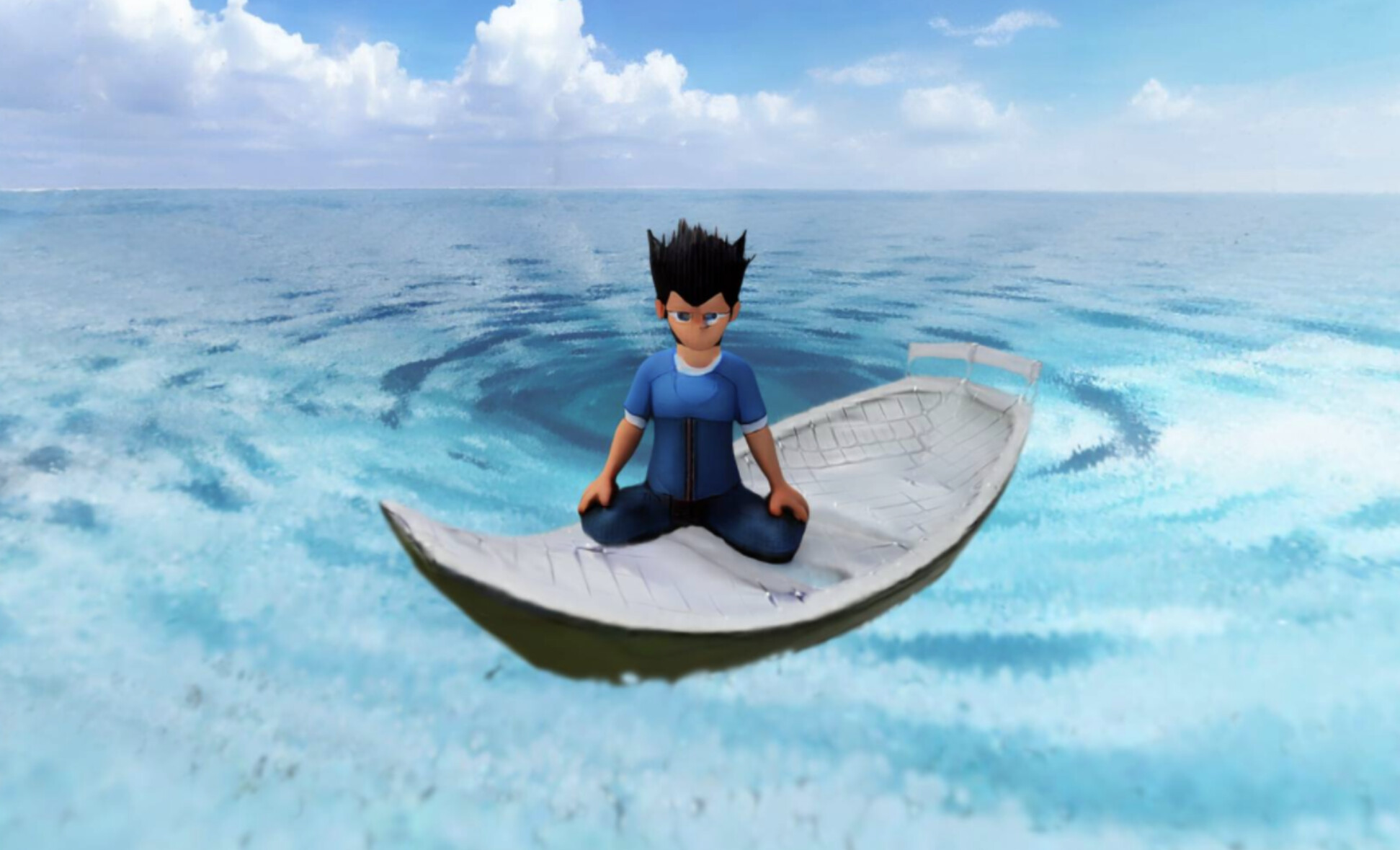} \\
  \rotatebox{90}{\footnotesize{\hspace{0.35em}WonderWorld}}
      & \includegraphics[width=0.3\linewidth]{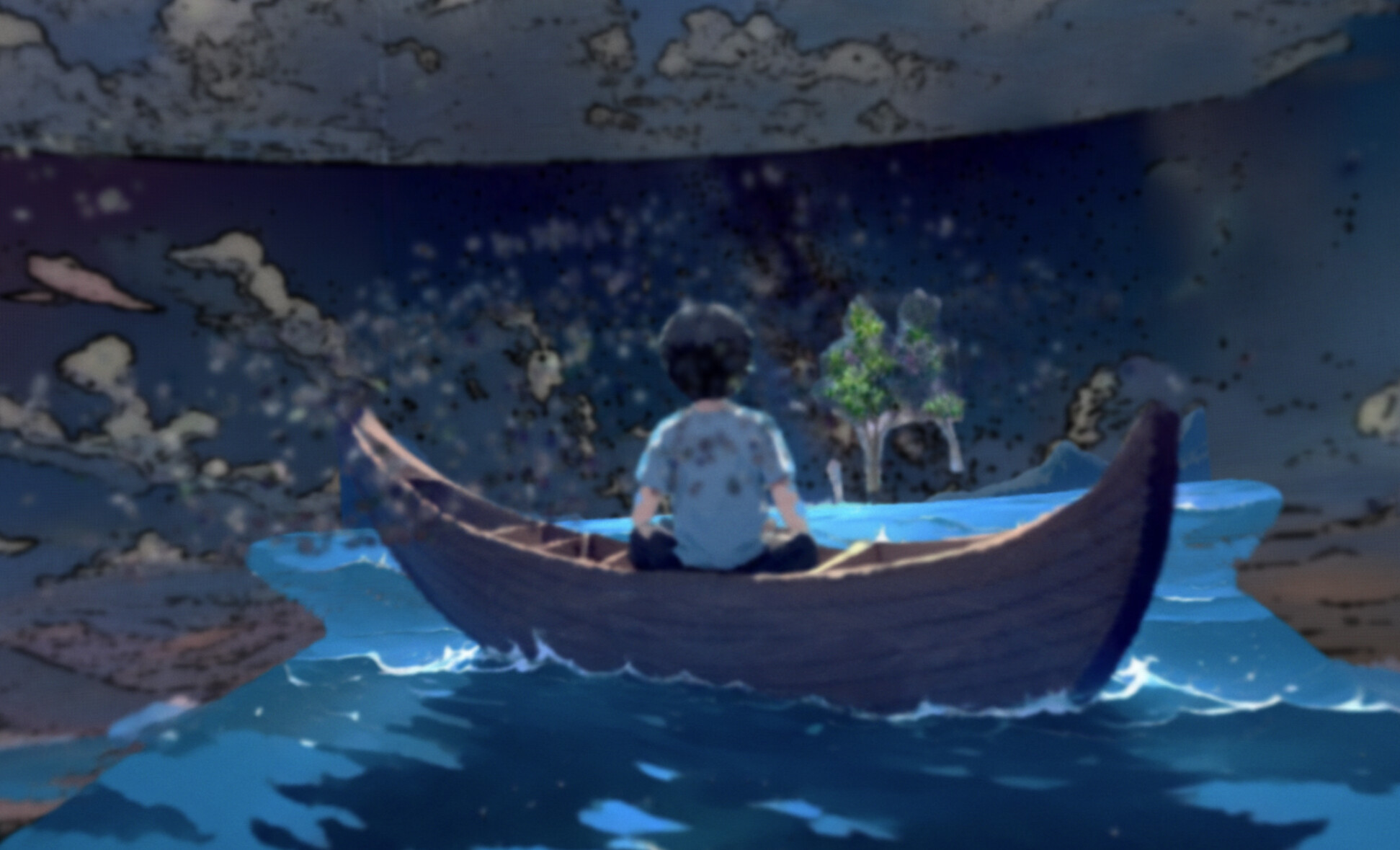}
      & \includegraphics[width=0.3\linewidth]{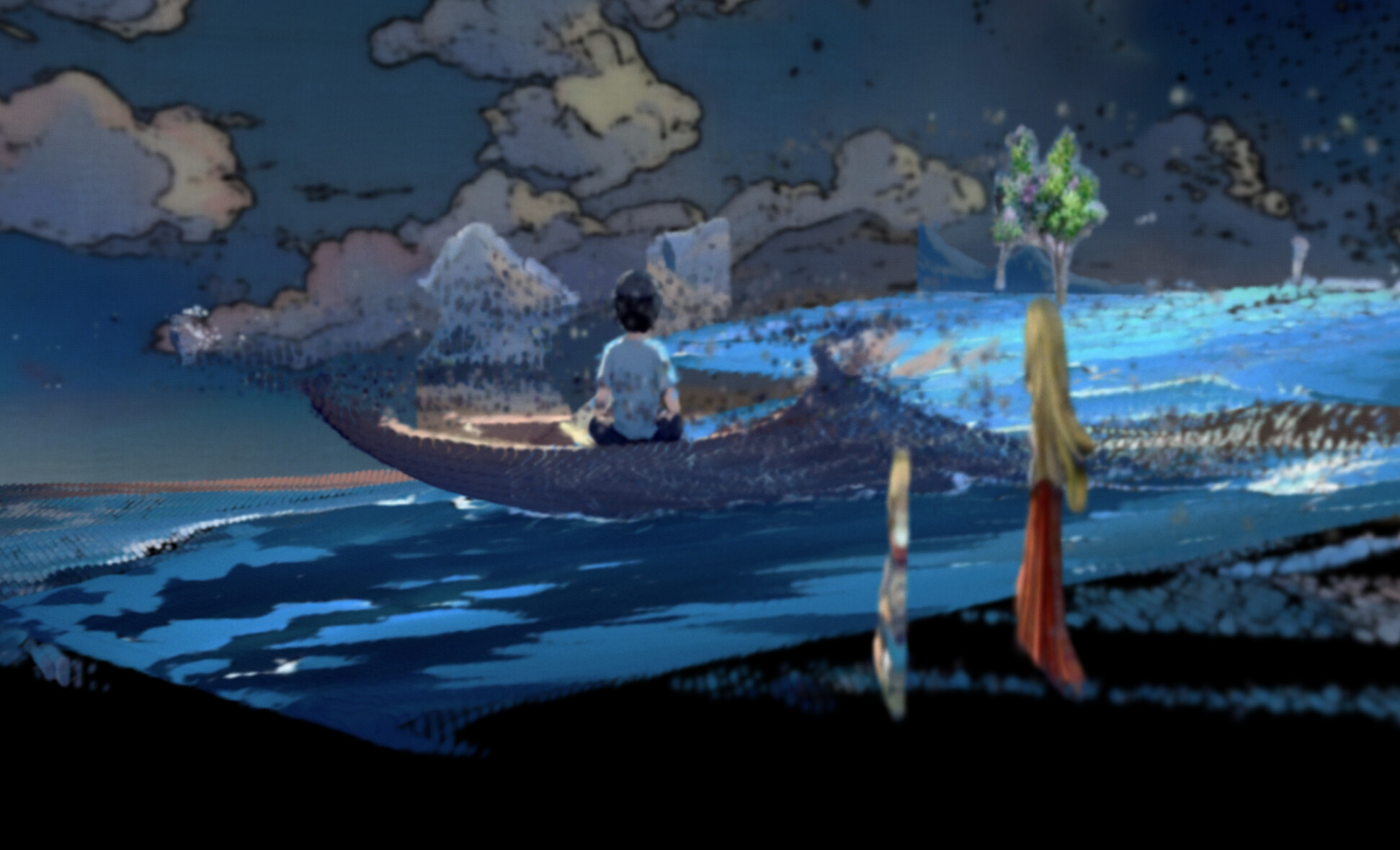}
      & \includegraphics[width=0.3\linewidth]{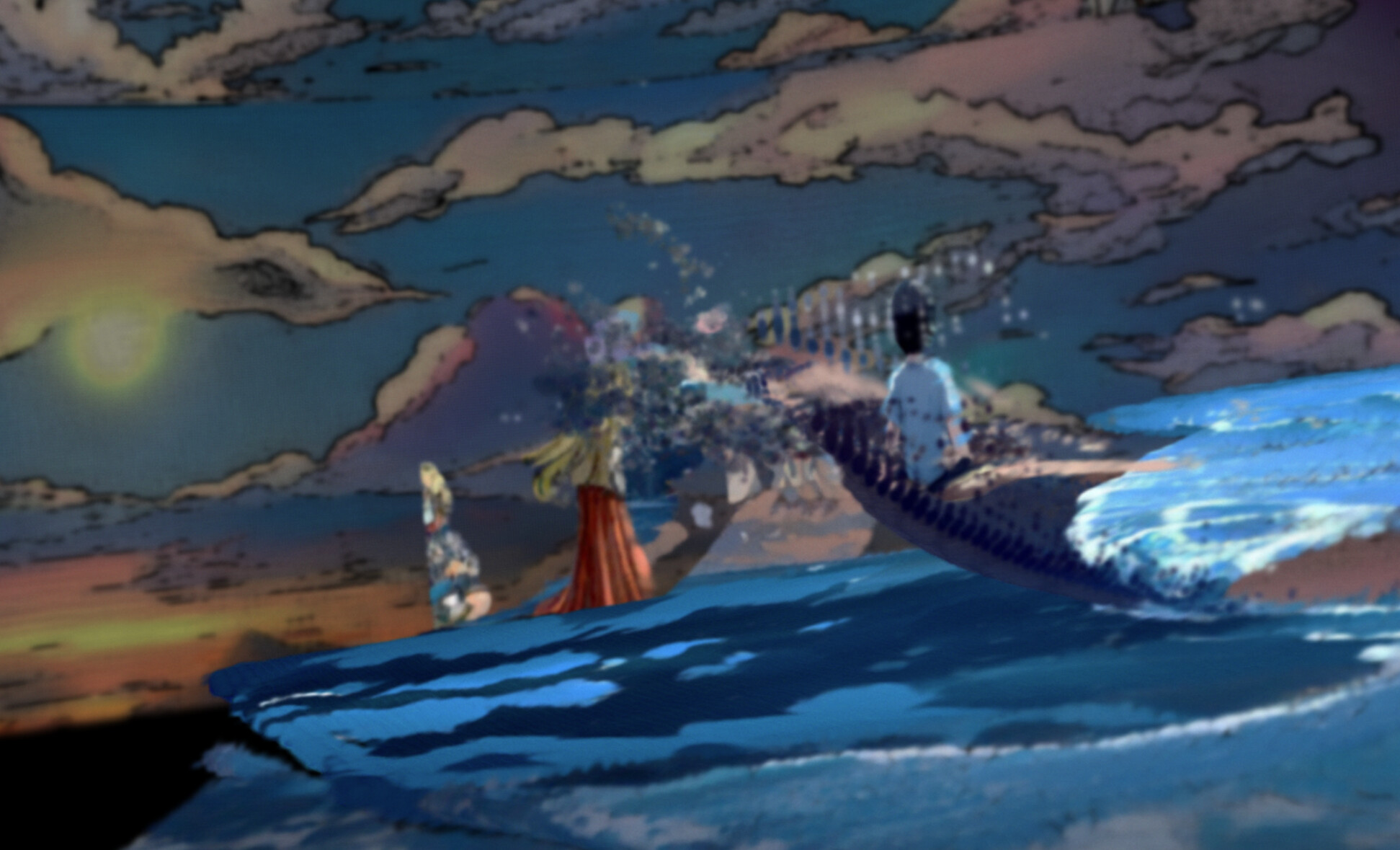} \\
  \rotatebox{90}{\footnotesize{\hspace{0.52em}DreamScene}}
      & \includegraphics[width=0.3\linewidth]{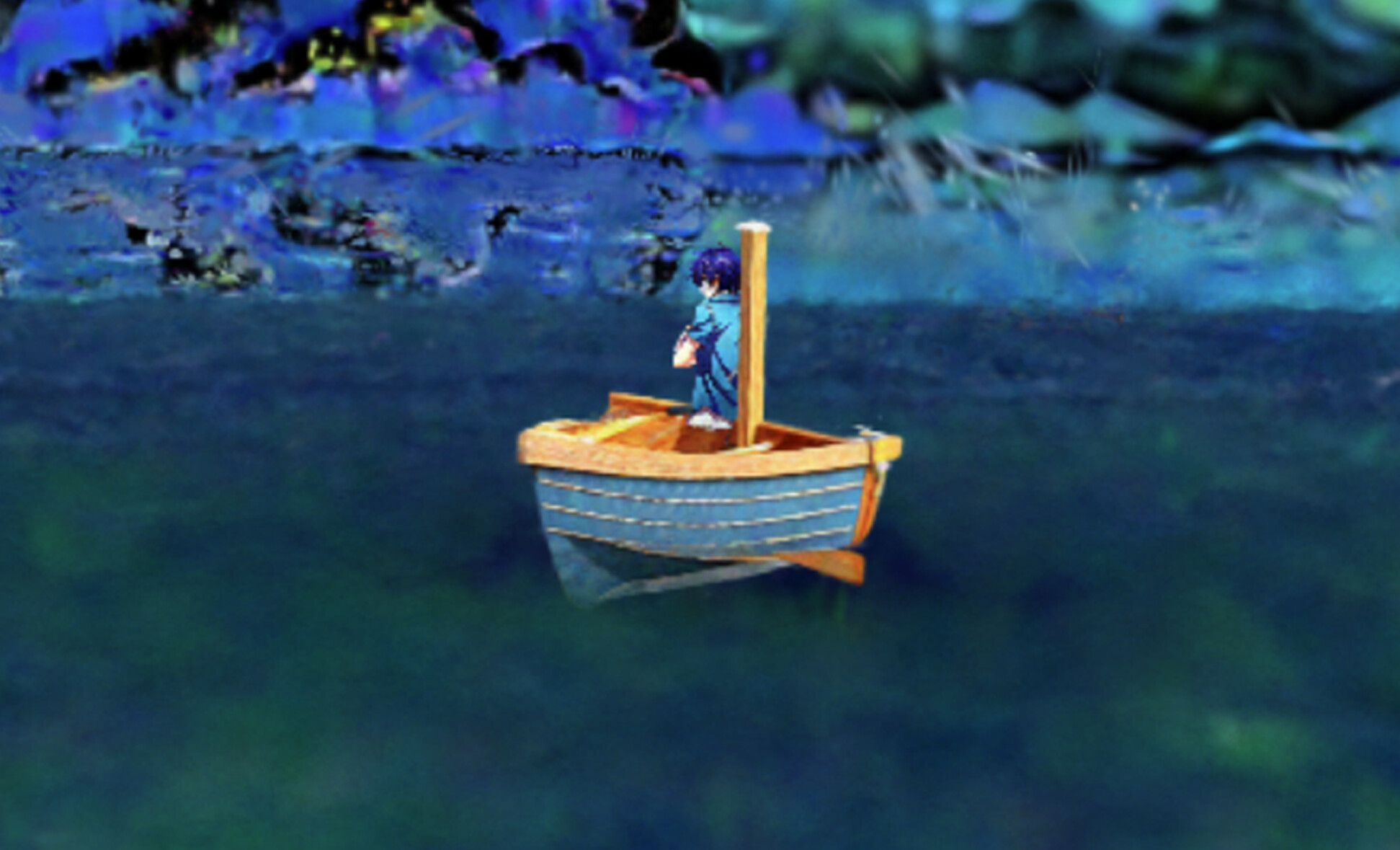}
      & \includegraphics[width=0.3\linewidth]{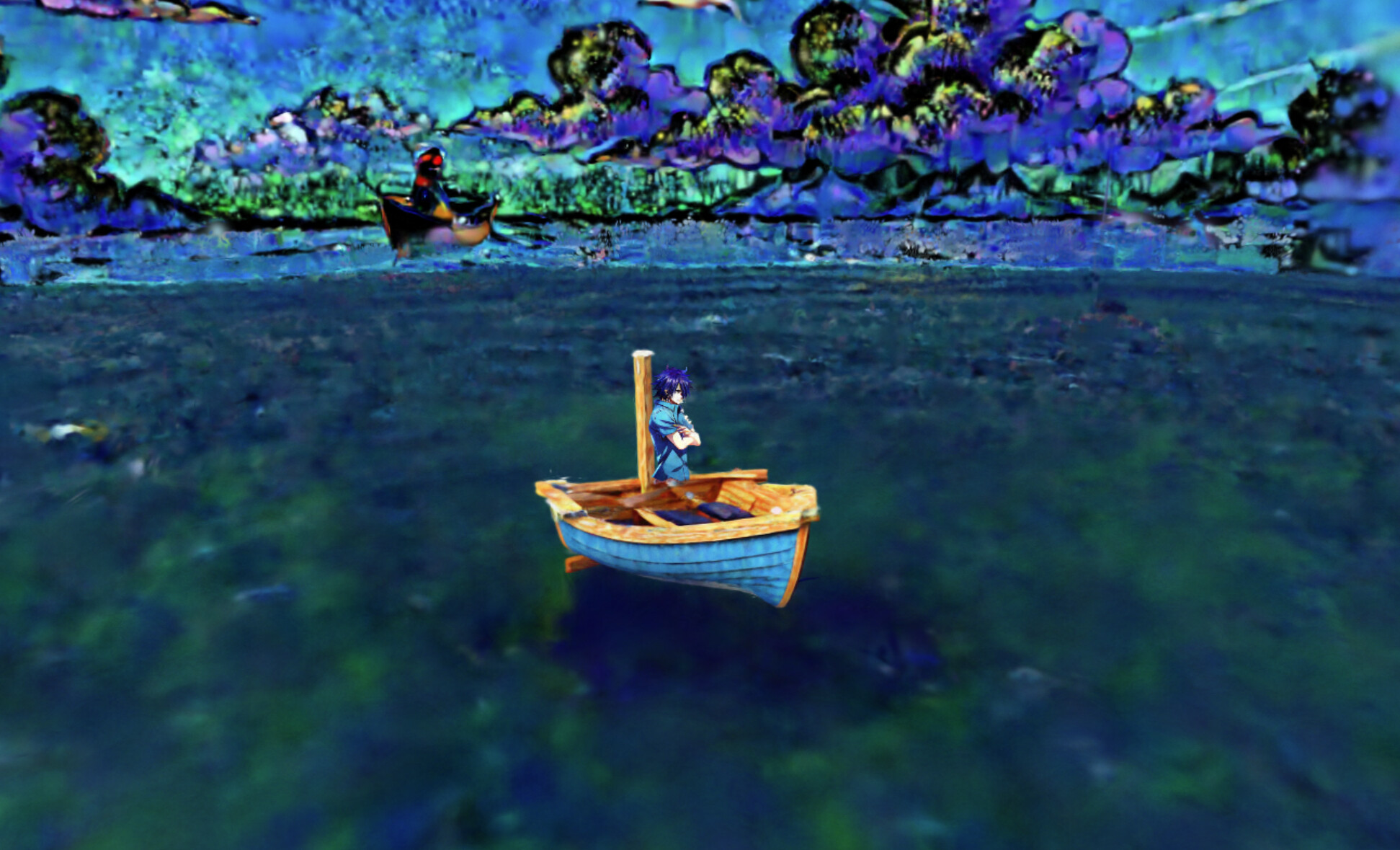}
      & \includegraphics[width=0.3\linewidth]{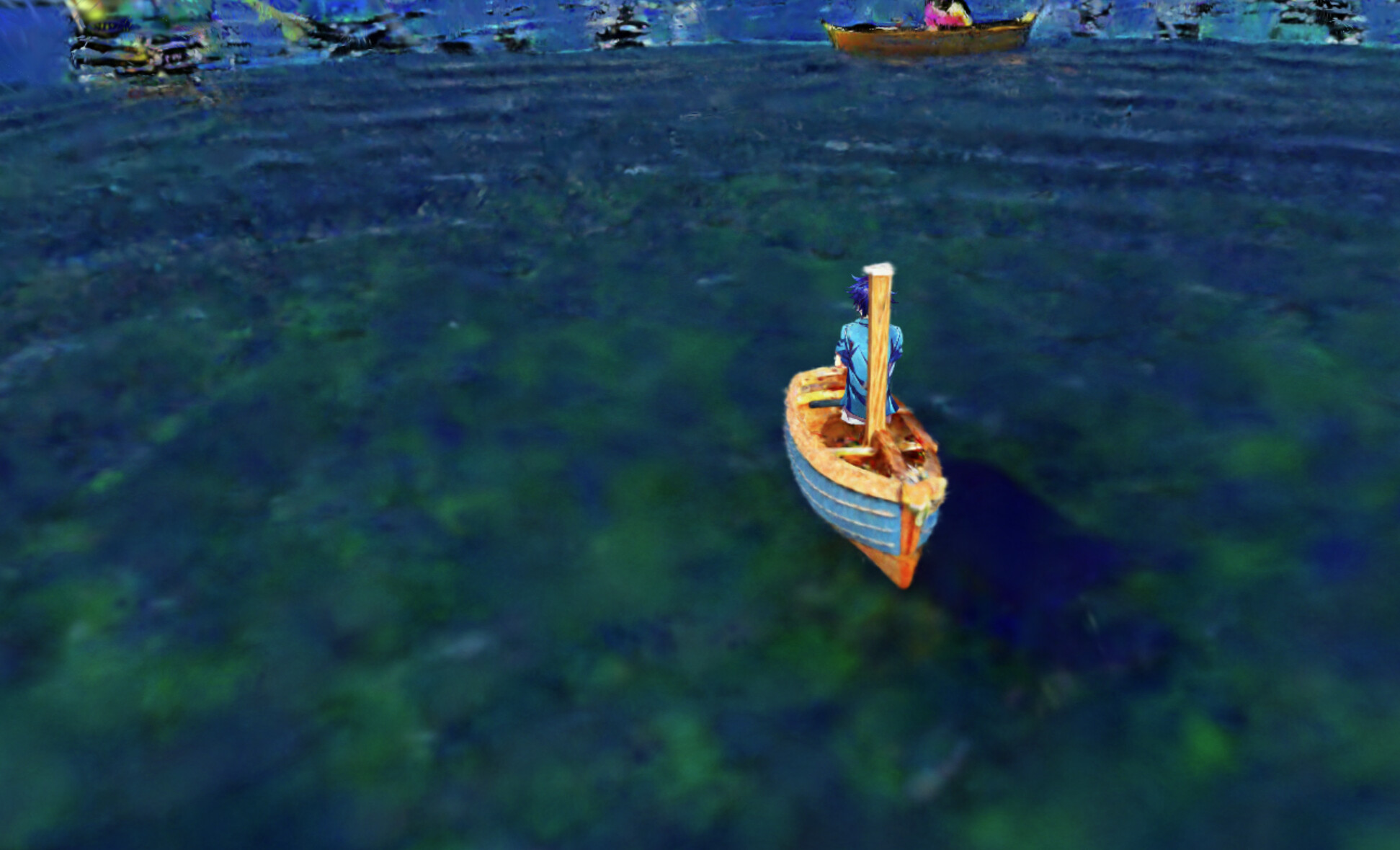} \\
\rotatebox{90}{\footnotesize{\hspace{0.4em}ViewCrafter}}
      & \includegraphics[width=0.3\linewidth]{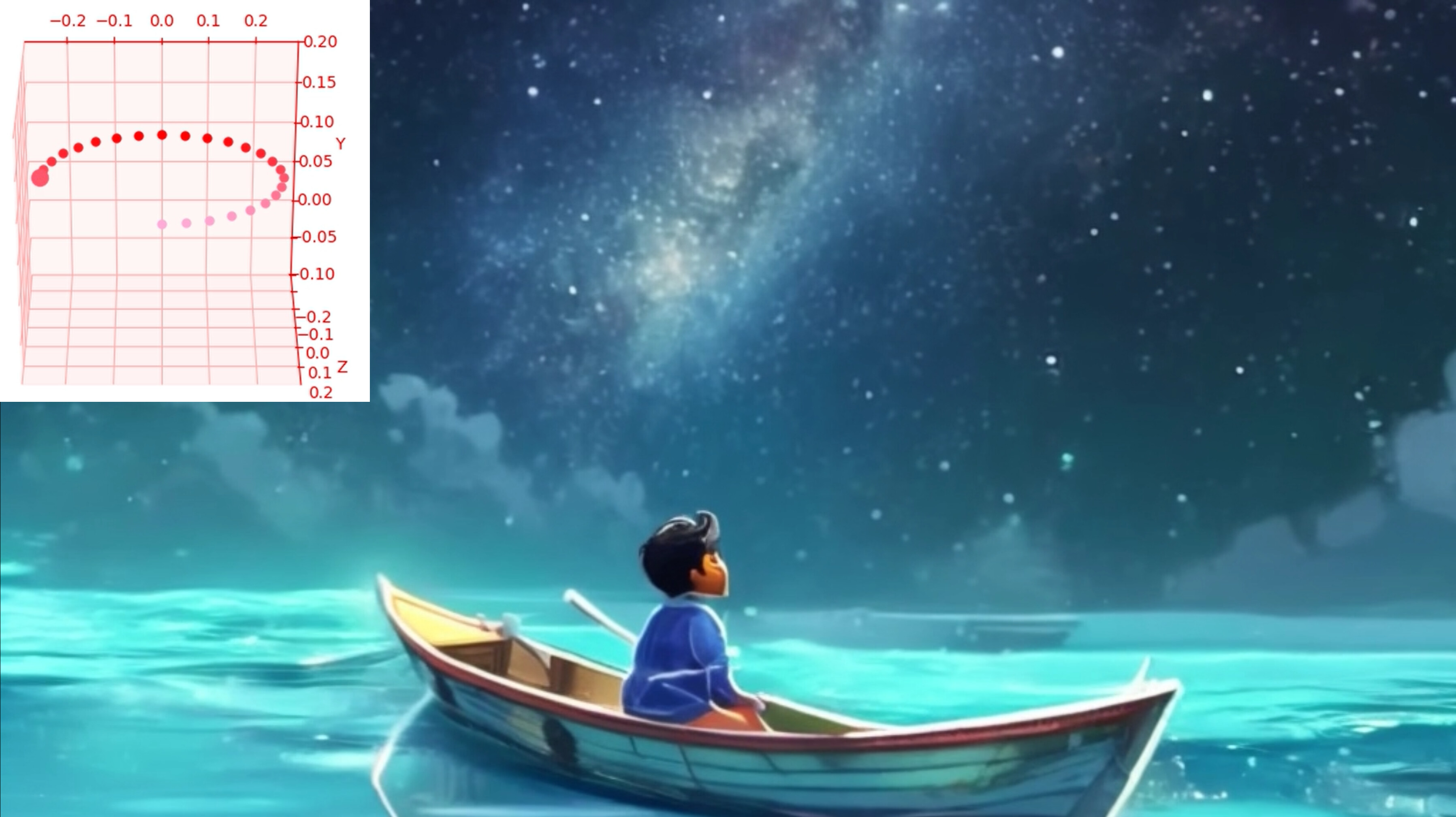}
      & \includegraphics[width=0.3\linewidth]{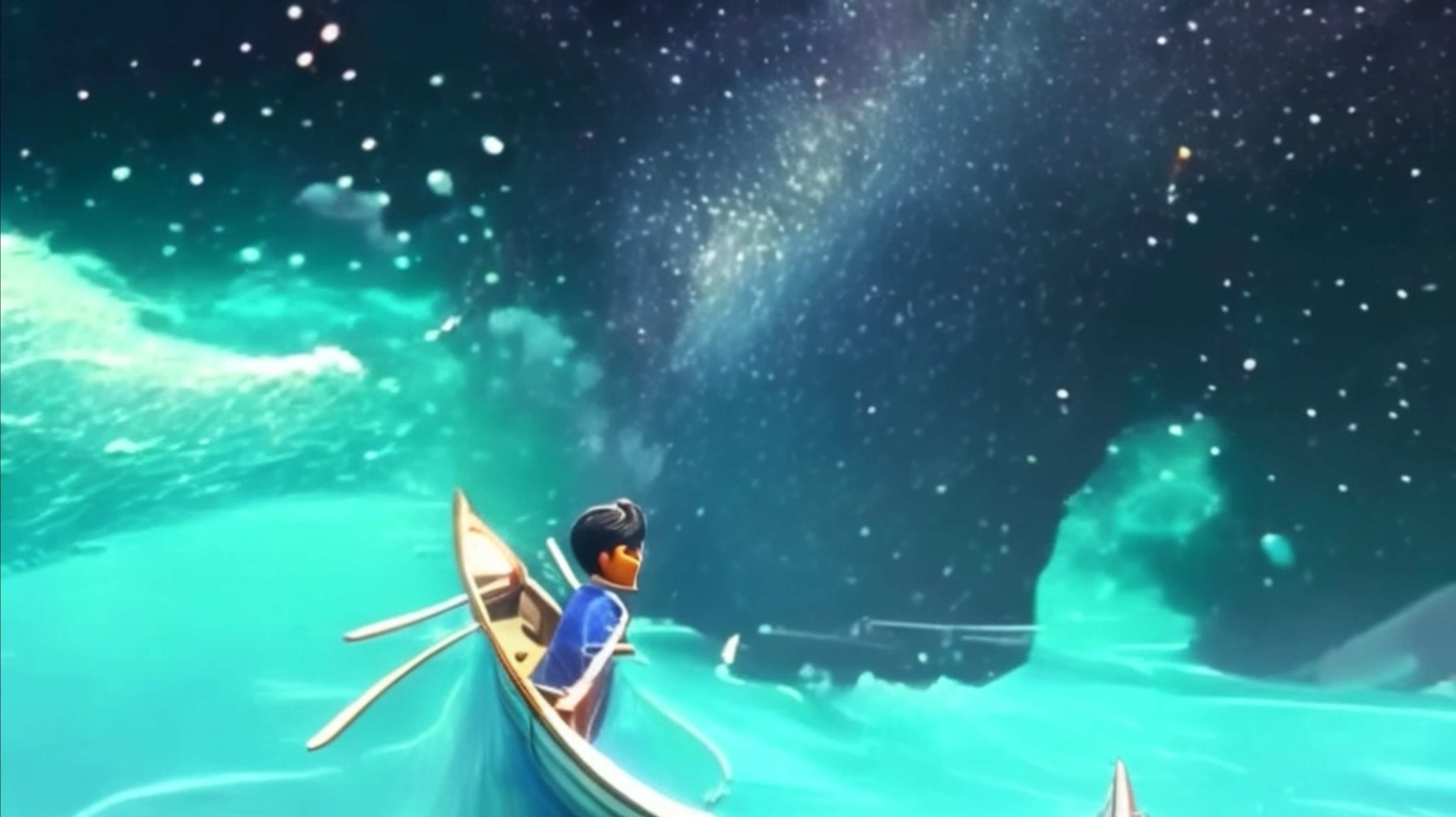}
      & \includegraphics[width=0.3\linewidth]{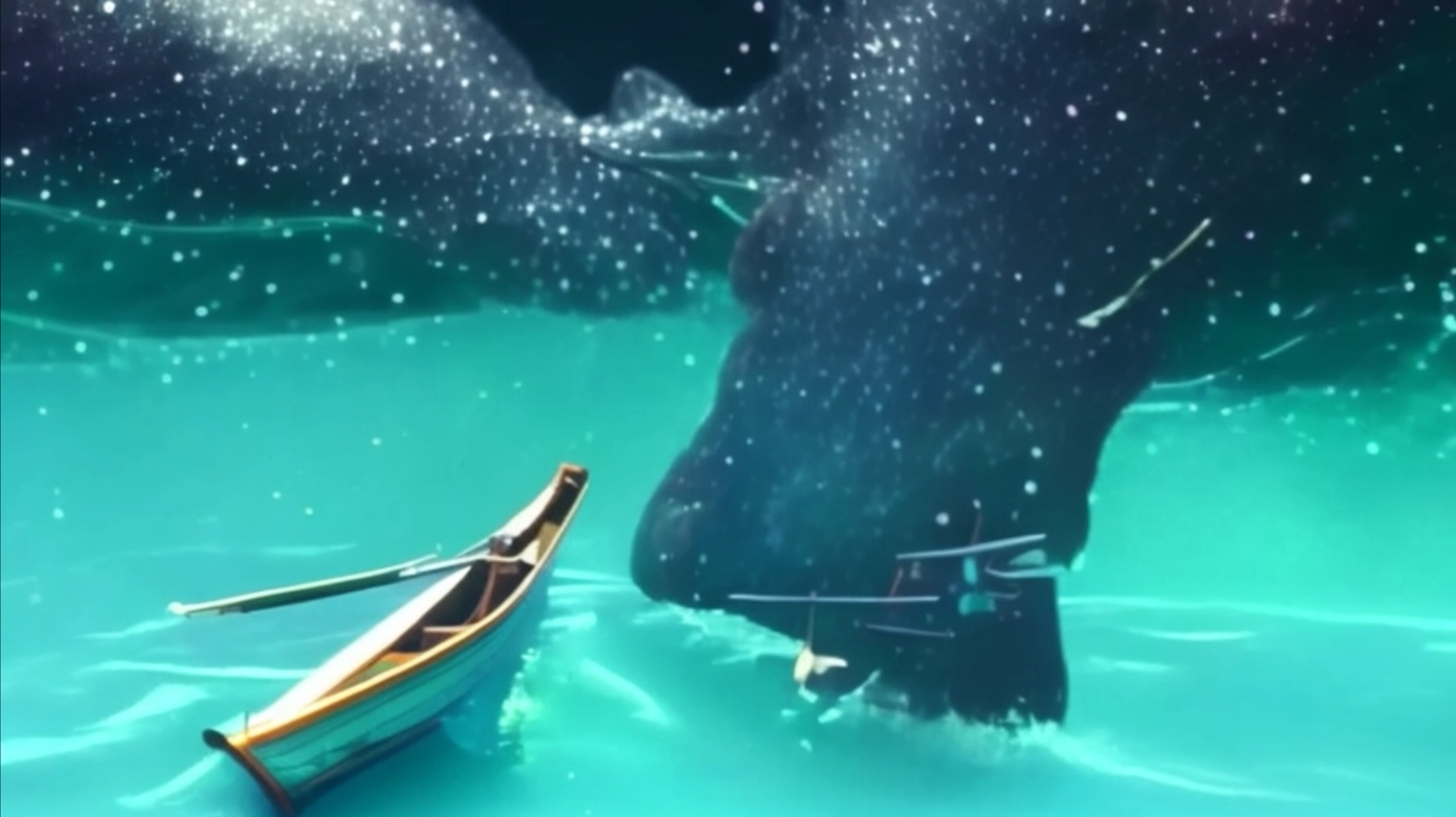}
  \end{tabular}
  \caption{ 
 Qualitative evaluation against WonderWorld~\cite{yu2024wonderworld}, DreamScene~\cite{dreamsceneLi2024} and ViewCrafter~\cite{yu2024viewcraftertamingvideodiffusion}. WonderWorld degenerates into flat, view-dependent renderings when shown from out-of-context viewpoints, whereas DreamScene exhibits reduced fidelity in complex regions and object poses while expecting the positioning of objects to be known a priori. ViewCrafter combines video diffusion and reconstruction models but it is not able to maintain 3D coherence for long camera paths. 
 }
  \label{fig:results_secondary}
\end{figure}

In this section we provide a quantitative and qualitative evaluation on the scene generation task, as well as an ablation of the core components of our pipeline. 
To assess the effectiveness of our approach in generating 3D scenes from text prompts, we focus on its ability to produce coherent and immersive scenes with high-quality and structurally consistent novel view synthesis.  
We compare our method against three state-of-the-art approaches for immersive scene generation: LayerPano3D~\cite{yang2024layerpano3d}, 
Text2Room~\cite{hoellein2023text2room}, 
and DreamScene360~\cite{zhou2025dreamscene360}.
Additionally, in \figref{fig:results_secondary} we provide a qualitative comparison with other methods that share a similar goal but would require substantial integration effort to directly compare against. 

Our method is evaluated on 17 text prompts commonly found in recent literature \cite{shriram2024realmdreamertextdriven3dscene, yang2024layerpano3d}. For each scene we recorded a 6-second walk-through video (at $1024 \times 1024$px resolution) along a predefined camera trajectory \textit{different} from the one used for the optimization described in \secref{subsec:background_generation}, covering large parts of the scene’s geometry as well as occluded surfaces.
Examples of video frames are shown in \figref{fig:quali_results_eval}. We generated all scenes on a single H100, with each scene taking around 15 mins to generate, given an average of 7.5 objects per scene. We include more results in the supplementary material.

\begin{figure}[t]
    \centering
        \includegraphics[width=\linewidth]{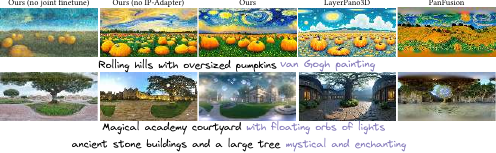}
    \caption{Qualitative ablation of out-of-domain sampling for panorama generator models. Our method captures the stylistic guidance faithfully, but without IP-Adapter it reverts back to its base panoramic distribution similarly to other recent methods. Simply using a perspective IP-Adapter does not work because of the mismatching distributions at training time. } 
    \label{fig:ablation}
\end{figure}

\begin{figure}
    \centering
    \includegraphics[width=\linewidth]{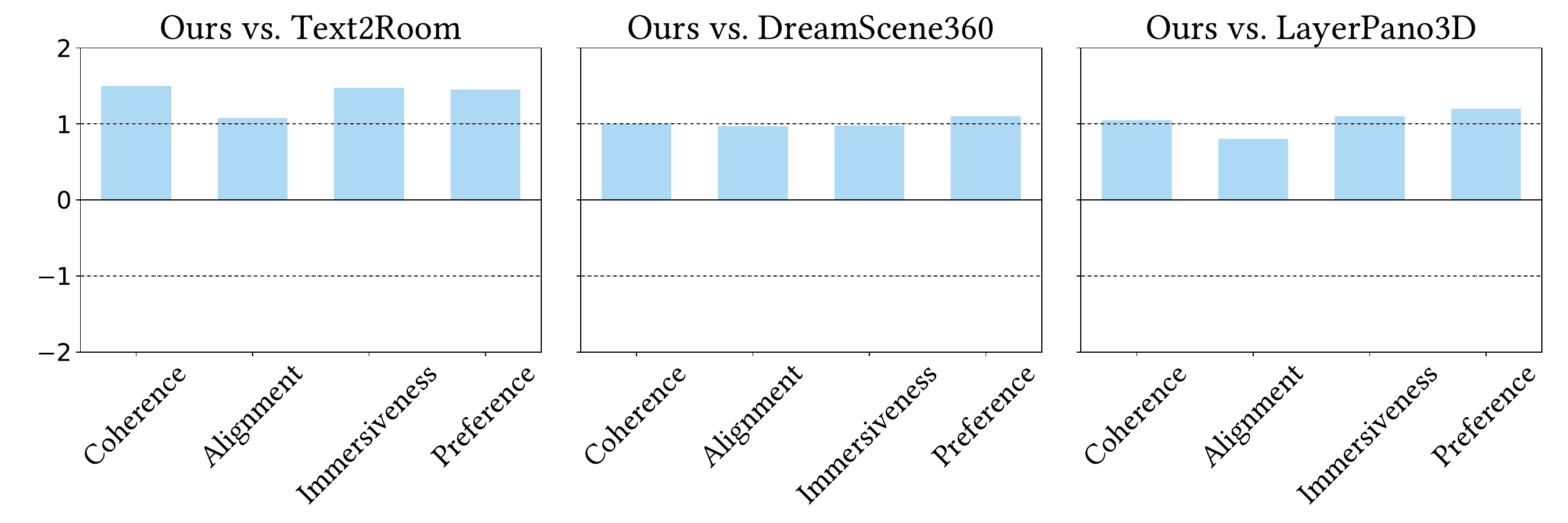}
    \caption{User study results for pairwise comparisons between our method, DreamScene360, Text2Room and LayerPano3D.
Ratings are given on a 5-point Likert scale capturing strong, weak and no preference between the approaches. The participants rated our approach higher than the comparison methods on all axes. }
    \label{fig:userstudy}
\end{figure}

\subsection{Quantitative Evaluation}

\begin{table}[t]
\centering
\caption{Quantitative evaluation of our method against Text2Room~\cite{hoellein2023text2room}, DreamScene360~\cite{zhou2025dreamscene360} and Layerpano3D~\cite{yang2024layerpano3d}. Metrics are averaged across all scenes and computed on 136 frames per scene. Our method has state-of-the-art performance in terms of perceptual image quality and visual aesthetics. The differences in CLIP scores (evaluated against the original text prompt) are small and should be interpreted with care (see main text).}
{\footnotesize
\setlength{\tabcolsep}{4pt}
\begin{tabular}{lcccc}
\toprule
Method & CLIP$\uparrow$ & Q-ALIGN$\uparrow$ & A-ALIGN$\uparrow$ & CLIP-IQA+$\uparrow$  \\
\midrule
Ours & 20.565 & \textbf{3.427} & \textbf{3.121} & \textbf{0.461}  \\
Text2Room & \textbf{20.779} & 2.998 & \underline{2.872} & 0.432 \\
DreamScene360 & 20.615 & 2.510 & 2.517 & 0.347 \\
Layerpano3D & \underline{20.721} & \underline{3.383} & 2.833 & \underline{0.421} \\
\bottomrule
\end{tabular}
}
\label{tab:quanteval}
\end{table}

\setlength{\tabcolsep}{2pt}
\begin{figure*}[t]
\includegraphics[width=\textwidth]{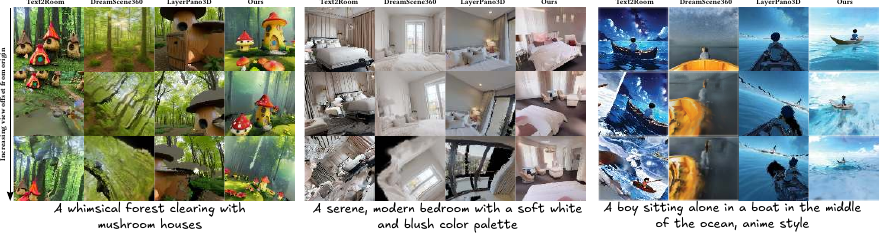}
\caption{
Qualitative comparison with Text2Room~\cite{hoellein2023text2room}, DreamScene360~\cite{zhou2025dreamscene360} and LayerPano3D~\cite{yang2024layerpano3d}.
Whereas scenes from these competing method suffer from disturbing artifacts for significant viewpoint changes, our method achieves coherent novel view synthesis for virtually all viewpoints.}
\label{fig:quali_results_eval}
\end{figure*}

\begin{figure}
    \centering
    \includegraphics[width=\linewidth]{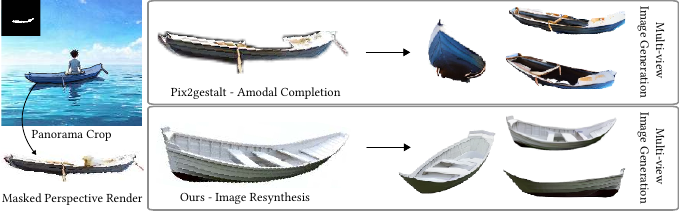}
    \caption{Diffusion-based amodal completion methods (such as pix2gestalt~\cite{ozguroglu2024pix2gestalt}) face challenges in producing high-quality inputs for multi-view image generation.
    Our method prioritizes structural fidelity by resynthesizing the reference image to yield higher quality 3D reconstructions.} 
    \label{fig:ablation_resynthesis}
\end{figure}

\begin{table}[t]
\centering
\caption{Ablation of the perspective conditioned IP-adapter for panorama generation. Out-of-domain prompt following improves without compromising generation quality. FID, KID and IS are evaluated on the test split of Matterport3d \cite{Matterport3D}. The relative difference in distribution similarity is significantly less relevant than CLIP improvement. }
{\footnotesize
\setlength{\tabcolsep}{4pt}
\begin{tabular}{lccccc}
\toprule
Method & CLIP $\uparrow$ & CLIP-persp $\uparrow$ & FID$\downarrow$ & KID $\downarrow$ & IS $\uparrow$ \\
\midrule
Full model           & \textbf{26.49} & \textbf{25.18} & \textbf{73.09} & \textbf{2.04} & \textbf{7.51} \\
w/o IP-adapter       & 24.82 & 24.10 & 74.34 & 2.40 & 7.14 \\
w/o joint fine-tuning & 17.54 & 14.48 & 437.67 & 33.91 & 2.98    \\
\bottomrule
\end{tabular}
}
\label{tab:quantablatepano}
\end{table}


We quantitatively evaluate prompt alignment and image quality.
To measure image quality and aesthetics, we use CLIP-IQA+~\cite{clipiqa}, Q-Align and A-Align~\cite{qalign} metrics. 
To quantify prompt alignment we compute the CLIPScore~\cite{clipscore} between the prompt and each rendered frame.
The results in \tabref{tab:quanteval} demonstrate that our method outperforms competing work in terms of both image quality and aesthetics metrics.
For completeness, we also evaluate CLIP scores on single views against the scene's text prompt. However, individual views typically capture only a small portion of the scene, making CLIP scores a less reliable metric for our novel view evaluation.

\figref{fig:quali_results_eval} provides a side-by-side comparison of the scenes generated by the different methods. In general, the results from Text2Room exhibit repetitive content and resemble stitched two-dimensional paintings---particularly in outdoor scenes---which limits the creation of immersive environments. However, the resulting images naturally excel in sharpness and detail, due to the method's use of meshes for 3D representation. 
LayerPano3D and DreamScene360 demonstrate some capability for novel view synthesis, but struggle when scenes are viewed with significant offsets, such as sideway perspectives of objects.

\subsection{User Study}
To assess the effectiveness of our model in generating immersive and navigable 3D scenes, we conducted a user study with 28 participants, including graduate students and professionals in graphics and vision. Participants evaluated short videos of scenes on: \textit{Coherence}: Do the spatial layout, object placement, and interactions in the scene form a logical and believable whole?; \textit{Compatibility}: Do the objects, materials, lighting, and textures align well with the given text prompt?; \textit{Immersiveness}: Does the scene convincingly represent a 3D environment, evoking a sense of depth, realism and presence?; \textit{Overall Preference}. Each comparison was rated from $-2$ to $+2$. 
As shown in \autoref{fig:userstudy}, our method was consistently preferred across all metrics and in particular was praised for its geometric structure in novel view synthesis, as shown by the \textit{coherence} and \textit{immersiveness} scores. The results were statistically significant ($p < .001$) according to the Wilcoxon signed-rank tests.

\subsubsection{Ablation}
We present a quantitative ablation of the panorama generator in Table~\ref{tab:quantablatepano}, highlighting the need for joint fine-tuning as well as the substantial prompt adherence obtained with the introduction of the IP-Adapter mechanism. A qualitative comparison with \cite{yang2024layerpano3d} and \cite{panfusion2024} is shown in \figref{fig:ablation}, noting that each model is built on top of different foundation models. Despite relying on only a subset of the public data used by \cite{yang2024layerpano3d}, our approach allows for more style-adhering images.\\
Our object generation module effectively addresses several common challenges in object extraction from images, such as panoramic distortion, low-resolution crops, and imperfect segmentation masks. Examples of these issues are illustrated in \figref{fig:ablation_resynthesis}. One alternative is to apply super-resolution and amodal completion but this approach typically requires multiple specialized modules, each with its own issues (\figref{fig:ablation_object}). Generating the same level of detail with super-resolution techniques on the panoramic image is still a challenge.

\vspace{-0.1em}

\section{Conclusion}
This work presented DreamAnywhere, a novel framework for generating immersive 3D scenes from text prompts by leveraging 360\degree{} images as foundational priors. Its modular, object-centric pipeline separates background and object generation, enabling high-fidelity object reconstruction and coherent scene completion through hybrid inpainting techniques. The modular nature of our pipeline has the disadvantage of having multiple possible points of failure. Conversely, this very modularity, serves as a significant advantage for targeted user intervention, which is \textit{often desired in generative applications.} It also facilitates replacement of individual modules (e.g. the image generator method with \cite{yang2024layerpano3d, text2lightchen2022, kalischek2025cubediffrepurposingdiffusionbasedimage}, the 3D inpainting step with \cite{yu2024viewcraftertamingvideodiffusion} or the object reconstruction pipeline with \cite{yao2025cast}). Compared to existing methods, our framework significantly enhances novel view synthesis coherence and offers robust performance for navigable 3D environments, as validated by quantitative results and strong user preference. This approach facilitates downstream applications like scene editing and simulation, paving the way for more scalable and interactive 3D content creation.

{
    \small
    \bibliographystyle{ieeenat_fullname}
    \bibliography{bibliography}

@String{tog = "ACM Transactions on Graphics"}

@String{neurips = "Proceedings of the Conference on Neural Information Processing Systems (NeurIPS)"}

@String{iclr = "Proceedings of the International Conference on Learning Representations (ICLR)"}

@String{cvpr = "Proceedings of the IEEE/CVF Conference on Computer Vision and Pattern Recognition (CVPR)"}

@String{cvprw = "Proceedings of the IEEE/CVF Conference on Computer Vision and Pattern Recognition Workshops (CVPRW)"}

@String{iccv = "Proceedings of the IEEE/CVF International Conference on Computer Vision (ICCV)"}

@String{eccv = "Proceedings of the European Conference on Computer Vision (ECCV)"}

@String{icml = "Proceedings of the International Conference on Machine Learning (ICML)"}

@String{threedv = "Proceedings of the International Conference on 3D Vision (3DV)"}

@inproceedings{lighting0,
  title={Shadowdiffusion: When Degradation Prior Meets Diffusion Model for Shadow Removal},
  author={Guo, Lanqing and Wang, Chong and Yang, Wenhan and Huang, Siyu and Wang, Yufei and Pfister, Hanspeter and Wen, Bihan},
  booktitle=cvpr,
  pages={14049--14058},
  year={2023}, 
  doi={10.1109/CVPR52729.2023.01350}
}

@article{voyager,
  title={Voyager: Long-Range and World-Consistent Video Diffusion for Explorable 3D Scene Generation},
  author={Huang, Tianyu and Zheng, Wangguandong and Wang, Tengfei and Liu, Yuhao and Wang, Zhenwei and Wu, Junta and Jiang, Jie and Li, Hui and Lau, Rynson WH and Zuo, Wangmeng and others},
  journal={arXiv preprint arXiv:2506.04225},
  year={2025}
}

@misc{liu2024infusion,
  title={InFusion: Inpainting 3D Gaussians via Learning Depth Completion from Diffusion Prior},
  author={Liu, Zhiheng and Ouyang, Hao and Wang, Qiuyu and Cheng, Ka Leong and Xiao, Jie and Zhu, Kai and Xue, Nan and Liu, Yu and Shen, Yujun and Cao, Yang},
  archivePrefix={arXiv},
  eprint={2404.11613},
  year={2024}
}

@misc{cosmos,
  title={Cosmos-Transfer1: Conditional World Generation with Adaptive Multimodal Control},
  author={Alhaija, Hassan Abu and Alvarez, Jose and Bala, Maciej and Cai, Tiffany and Cao, Tianshi and Cha, Liz and Chen, Joshua and Chen, Mike and Ferroni, Francesco and Fidler, Sanja and others},
  archivePrefix={arXiv},
  eprint={2503.14492},
  year={2025}
}

@inproceedings{lighting1,
  title={On The Detection of Synthetic Images Generated by Diffusion Models},
  author={Corvi, Riccardo and Cozzolino, Davide and Zingarini, Giada and Poggi, Giovanni and Nagano, Koki and Verdoliva, Luisa},
  booktitle={ICASSP 2023 IEEE International Conference on Acoustics, Speech and Signal Processing (ICASSP)},
  pages={1--5},
  year={2023},
doi={10.1109/ICASSP49357.2023.10095167}
}

@misc{surveyscene20254,
  title={Recent Advance in 3D Object and Scene Generation: A Survey},
  author={Tang, Xiang and Li, Ruotong and Fan, Xiaopeng},
  archivePrefix={arXiv},
  eprint={2504.11734},
  year={2025}
}

@misc{surveygen2023,
  title={Generative AI Meets 3D: A Survey on Text-to-3D in AIGC Era},
  author={Li, Chenghao and Zhang, Chaoning and Cho, Joseph and Waghwase, Atish and Lee, Lik-Hang and Rameau, Francois and Yang, Yang and Bae, Sung-Ho and Hong, Choong Seon},
  archivePrefix={arXiv},
  eprint={2305.06131},
  year={2023}
}

@inproceedings{kheradmand2024mcmc,
    title = {{3D Gaussian Splatting as Markov Chain Monte Carlo}},
    author = {Kheradmand, Shakiba and Rebain, Daniel and Sharma, Gopal and Sun, Weiwei and Tseng, Yang-Che and Isack, Hossam and Kar, Abhishek and Tagliasacchi, Andrea and Yi, Kwang Moo},
    booktitle = neurips,
    year = {2024},
    url={https://openreview.net/pdf?id=UCSt4gk6iX}
   }

@misc{huang2024midi,
  title={Midi: Multi-instance diffusion for single image to 3d scene generation},
  author={Huang, Zehuan and Guo, Yuan-Chen and An, Xingqiao and Yang, Yunhan and Li, Yangguang and Zou, Zi-Xin and Liang, Ding and Liu, Xihui and Cao, Yan-Pei and Sheng, Lu},
  booktitle={Proceedings of the Computer Vision and Pattern Recognition Conference},
  pages={23646--23657},
  year={2025}
}

@misc{Dogaru2024Gen3DSR,
        title={Generalizable 3D Scene Reconstruction via Divide and Conquer from a Single View},
        author={Ardelean, Andreea and Özer, Mert and Egger, Bernhard},
        booktitle = {International Conference on 3D Vision (3DV)},
        year={2025}
    }

@InProceedings{hoellein2023text2room,
    author    = {H\"ollein, Lukas and Cao, Ang and Owens, Andrew and Johnson, Justin and Nie{\ss}ner, Matthias},
    title     = {Text2Room: Extracting Textured 3D Meshes from 2D Text-to-Image Models},
    booktitle = iccv,
    month     = {October},
    year      = {2023},
    pages     = {7909-7920},
    doi       = {10.1109/ICCV51070.2023.00727},
}

@inproceedings{schult24controlroom3d,
  author    = {Schult, Jonas and Tsai, Sam and H\"ollein, Lukas and Wu, Bichen and Wang, Jialiang and Ma, Chih-Yao and Li, Kunpeng and Wang, Xiaofang and Wimbauer, Felix and He, Zijian and Zhang, Peizhao and Leibe, Bastian and Vajda, Peter and Hou, Ji},
  title     = {ControlRoom3D: Room Generation using Semantic Proxy Rooms},
  booktitle = cvpr,
  year      = {2024},
doi = {10.1109/CVPR52733.2024.00593},
pages = {6201-6210}
}

@misc{shriram2024realmdreamertextdriven3dscene,
      title={RealmDreamer: Text-Driven 3D Scene Generation with Inpainting and Depth Diffusion}, 
      author={Jaidev Shriram and Alex Trevithick and Lingjie Liu and Ravi Ramamoorthi},
      year={2024},
      eprint={2404.07199},
      archivePrefix={arXiv},
}

@inproceedings{yang2024scenecraft,
        title={SceneCraft: Layout-Guided 3D Scene Generation},
        author={Yang, Xiuyu and Man, Yunze and Chen, Jun-Kun and Wang, Yu-Xiong},
        booktitle=neurips,
        year={2024},
        url={https://openreview.net/pdf?id=CTvxvAcSJN}
      }

@misc{yu2024wonderworld,
  title={Wonderworld: Interactive 3d scene generation from a single image},
  author={Yu, Hong-Xing and Duan, Haoyi and Herrmann, Charles and Freeman, William T and Wu, Jiajun},
  booktitle={Proceedings of the Computer Vision and Pattern Recognition Conference},
  pages={5916--5926},
  year={2025}
}

@inproceedings{yu2023wonderjourney,
    title={WonderJourney: Going from Anywhere to Everywhere},
    author={Hong-Xing Yu and Haoyi Duan and Junhwa Hur and Kyle Sargent and Michael Rubinstein and William T. Freeman and Forrester Cole and Deqing Sun and Noah Snavely and Jiajun Wu and Charles Herrmann},
  booktitle = cvpr,
    doi={10.1109/CVPR52733.2024.00636},
    year={2024}
}

@inproceedings{zhang2023scenewiz3d,
              author = {Qihang Zhang and Chaoyang Wang and Aliaksandr Siarohin and Peiye Zhuang and Yinghao Xu and Ceyuan Yang and Dahua Lin and Bo Dai and Bolei Zhou and Sergey Tulyakov and Hsin-Ying Lee},
              title = {{SceneWiz3D}: Towards Text-guided {3D} Scene Composition},
              booktitle = cvpr,
                url={https://openreview.net/pdf?id=a5C3JmS4S5},
              year = {2023}
}

@inproceedings{matterport2024defurnishing,
    author    = {Slavcheva, Mira and Gausebeck, Dave and Chen, Kevin and Buchhofer, David and Sabik, Azwad and Ma, Chen and Dhillon, Sachal and Brandt, Olaf and Dolhasz, Alan},
    title     = {An Empty Room is All We Want: Automatic Defurnishing of Indoor Panoramas},
    booktitle = cvprw,
    year      = {2024},
doi = {10.1109/CVPRW63382.2024.00734},
pages = {7384-7394}
}

@inproceedings{poole2023dreamfusion,
title={{DreamFusion: Text-to-3D using 2D Diffusion}},
author={Ben Poole and Ajay Jain and Jonathan T. Barron and Ben Mildenhall},
booktitle=iclr,
year={2023},
url={https://openreview.net/forum?id=FjNys5c7VyY}
}

@misc{shi2023zero123plus,
      title={Zero123++: a Single Image to Consistent Multi-view Diffusion Base Model}, 
      author={Ruoxi Shi and Hansheng Chen and Zhuoyang Zhang and Minghua Liu and Chao Xu and Xinyue Wei and Linghao Chen and Chong Zeng and Hao Su},
      year={2023},
      eprint={2310.15110},
      archivePrefix={arXiv}
}

@misc{xu2024instantmesh,
  title={InstantMesh: Efficient 3D Mesh Generation from a Single Image with Sparse-view Large Reconstruction Models},
  author={Xu, Jiale and Cheng, Weihao and Gao, Yiming and Wang, Xintao and Gao, Shenghua and Shan, Ying},
  eprint={2404.07191},
  archivePrefix={arXiv},
  year={2024}
}

@inproceedings{wu2023panodiffusion,
  title={PanoDiffusion: 360-degree Panorama Outpainting via Diffusion},
  author={Wu, Tianhao and Zheng, Chuanxia and Cham, Tat-Jen},
  booktitle=iclr,
  year={2023},
  url={https://openreview.net/forum?id=ZNzDXDFZ0B},
}

@inproceedings{depreyarea2021360monodepth,
	title={{360MonoDepth}: High-Resolution 360° Monocular Depth Estimation},
	author={Manuel Rey-Area and Mingze Yuan and Christian Richardt},
  booktitle=cvpr, 
  pages={3752-3762},
  doi={10.1109/CVPR52688.2022.00374},
 	year={2022}}

@inproceedings{zhou2025dreamscene360,
      title={DreamScene360: Unconstrained Text-to-3D Scene Generation with Panoramic Gaussian Splatting},
      author={Zhou, Shijie and Fan, Zhiwen and Xu, Dejia and Chang, Haoran and Chari, Pradyumna and Bharadwaj, Tejas and You, Suya and Wang, Zhangyang and Kadambi, Achuta},
      booktitle=eccv,
      pages={324--342},
      year={2025},
      doi={10.1007/978-3-031-72658-3_19},
}

@InProceedings{dreamsceneLi2024,
author="Li, Haoran
and Shi, Haolin
and Zhang, Wenli
and Wu, Wenjun
and Liao, Yong
and Wang, Lin
and Lee, Lik-Hang
and Zhou, Peng Yuan",
title="DreamScene: 3D Gaussian-Based Text-to-3D Scene Generation via Formation Pattern Sampling",
booktitle=eccv,
year="2025",
pages="214--230",
doi={https://doi.org/10.1007/978-3-031-72904-1_13}
}

@inproceedings{ye2024diffpano,
title={DiffPano: Scalable and Consistent Text to Panorama Generation with Spherical Epipolar-Aware Diffusion},
author={Weicai Ye and Chenhao Ji and Zheng Chen and Junyao Gao and Xiaoshui Huang and Song-Hai Zhang and Wanli Ouyang and Tong He and Cairong Zhao and Guofeng Zhang},
booktitle=neurips,
year={2024},
url={https://openreview.net/forum?id=YIOvR40hSo},
}

@INPROCEEDINGS{EGformer2023,
  author={Yun, Ilwi and Shin, Chanyong and Lee, Hyunku and Lee, Hyuk-Jae and Rhee, Chae Eun},
  booktitle=iccv, 
  title={EGformer: Equirectangular Geometry-biased Transformer for 360 Depth Estimation}, 
  year={2023},
  volume={},
  number={},
  pages={6078-6089},
  doi={10.1109/ICCV51070.2023.00561}
}

@inproceedings{zhou2024DeepPriorAssembly,
      title = {Zero-Shot Scene Reconstruction from Single Images with Deep Prior Assembly},
      author = {Zhou, Junsheng and Liu, Yu-Shen and Han, Zhizhong},
      booktitle = neurips,
      year = {2024}, 
      url={https://openreview.net/pdf?id=SoTK84ewb7}
  }

@misc{han2024reparocompositional3dassets,
      title={REPARO: Compositional 3D Assets Generation with Differentiable 3D Layout Alignment}, 
      author={Haonan Han and Rui Yang and Huan Liao and Jiankai Xing and Zunnan Xu and Xiaoming Yu and Junwei Zha and Xiu Li and Wanhua Li},
      year={2024},
      eprint={2405.18525},
      archivePrefix={arXiv},
}

@InProceedings{chen2025comboverse,
    author = {Chen, Yongwei and Wang, Tengfei and Wu, Tong and Pan, Xingang and Jia, Kui and Liu, Ziwei},
    title="ComboVerse: Compositional 3D Assets Creation Using Spatially-Aware Diffusion Guidance",
    booktitle=eccv,
    year="2025",
    pages="128--146",
    doi = {10.1007/978-3-031-72691-0_8},
}

@misc{paliwal2024panodreamer,
    title={PanoDreamer: 3D Panorama Synthesis from a Single Image},
    author={Avinash Paliwal and Xilong Zhou and Andrii Tsarov and Nima Khademi Kalantari},
    archivePrefix={arXiv},
    eprint={2412.04827},
    year={2024}
}

@misc{li2024scenedreamer360,
  title={Scenedreamer: Unbounded 3d scene generation from 2d image collections},
  author={Chen, Zhaoxi and Wang, Guangcong and Liu, Ziwei},
  journal={IEEE transactions on pattern analysis and machine intelligence},
  volume={45},
  number={12},
  pages={15562--15576},
  year={2023},
  publisher={IEEE}
}

@misc{zhou2024holodreamerholistic3dpanoramic,
      title={HoloDreamer: Holistic 3D Panoramic World Generation from Text Descriptions}, 
      author={Haiyang Zhou and Xinhua Cheng and Wangbo Yu and Yonghong Tian and Li Yuan},
      year={2024},
      eprint={2407.15187},
      archivePrefix={arXiv},
}

@inproceedings{ARAP,
author = {Olga Sorkine and Marc Alexa},
title = {As-Rigid-As-Possible Surface Modeling},
booktitle = {Proceedings of EUROGRAPHICS/ACM SIGGRAPH Symposium on Geometry Processing},
year = {2007},
pages = {109--116},
doi={10.5555/1281991.1282006}
}

@article{yang2024layerpano3d,
  title={Layerpano3d: Layered 3d panorama for hyper-immersive scene generation},
  author={Yang, Shuai and Tan, Jing and Zhang, Mengchen and Wu, Tong and Wetzstein, Gordon and Liu, Ziwei and Lin, Dahua},
  booktitle={Proceedings of the Special Interest Group on Computer Graphics and Interactive Techniques Conference Conference Papers},
  pages={1--10},
  year={2025}
}

@INPROCEEDINGS{Lama2022,
  author={Suvorov, Roman and Logacheva, Elizaveta and Mashikhin, Anton and Remizova, Anastasia and Ashukha, Arsenii and Silvestrov, Aleksei and Kong, Naejin and Goka, Harshith and Park, Kiwoong and Lempitsky, Victor},
  booktitle={Proceedings of the IEEE/CVF Winter Conference on Applications of Computer Vision (WACV)}, 
  title={Resolution-robust Large Mask Inpainting with Fourier Convolutions}, 
  year={2022},
  volume={},
  number={},
  pages={3172-3182},
  doi={10.1109/WACV51458.2022.00323}
}

@inproceedings{scenescape2023,
    author = {Fridman, Rafail and Abecasis, Amit and Kasten, Yoni and Dekel, Tali},
    title = {SceneScape: Text-Driven Consistent Scene Generation},
    year = {2024},
    booktitle = neurips,
    articleno = {1734},
    numpages = {18},
    doi={10.5555/3666122.3667856},
}

@inproceedings{mast3r_arxiv24,
      title={Grounding Image Matching in 3D with MASt3R}, 
      author={Vincent Leroy and Yohann Cabon and Jerome Revaud},
      year={2024},
      booktitle=eccv,
url={https://doi.org/10.1007/978-3-031-73220-1_5}
}

@inproceedings{epstein2024disentangled,
      title={Disentangled 3D Scene Generation with Layout Learning},
      author={Dave Epstein and Ben Poole and Ben Mildenhall and Alexei A. Efros and Aleksander Holynski},
      year={2024},
      booktitle=icml,
      doi={10.5555/3692070.3692570}
}

@article{yao2025cast,
  title={Cast: Component-aligned 3d scene reconstruction from an rgb image},
  author={Yao, Kaixin and Zhang, Longwen and Yan, Xinhao and Zeng, Yan and Zhang, Qixuan and Xu, Lan and Yang, Wei and Gu, Jiayuan and Yu, Jingyi},
  journal={ACM Transactions on Graphics (TOG)},
  volume={44},
  number={4},
  pages={1--19},
  year={2025},
  publisher={ACM New York, NY, USA}
}

@inproceedings{MVDiffusion2024,
author = {Tang, Shitao and Zhang, Fuyang and Chen, Jiacheng and Wang, Peng and Furukawa, Yasutaka},
title = {MVDiffusion: Enabling Holistic Multi-view Image Generation with Correspondence-Aware Diffusion},
year = {2024},
booktitle = neurips,
articleno = {2229},
numpages = {32},
url={https://openreview.net/pdf?id=vA0vj1mY77}
}

@misc{ye2023ip,
  title={IP-Adapter: Text Compatible Image Prompt Adapter for Text-to-Image Diffusion Models},
  author={Ye, Hu and Zhang, Jun and Liu, Sibo and Han, Xiao and Yang, Wei},
  eprint={2308.06721},
archivePrefix={arXiv},  
year={2023}
}

@inproceedings{panfusion2024,
  title={Taming Stable Diffusion for Text to 360\textdegree Panorama Image Generation},
  author={Zhang, Cheng and Wu, Qianyi and Cruz Gambardella, Camilo and Huang, Xiaoshui and Phung, Dinh and Ouyang, Wanli and Cai, Jianfei},
  booktitle=cvpr,
  year={2024},
  doi={10.1109/CVPR52733.2024.00607}
}

@article{Zhang2023Text2NeRFT3,
  title={Text2NeRF: Text-Driven 3D Scene Generation With Neural Radiance Fields},
  author={Jingbo Zhang and Xiaoyu Li and Ziyu Wan and Can Wang and Jing Liao},
  journal={IEEE Transactions on Visualization and Computer Graphics},
  year={2023},
  volume={30},
  pages={7749-7762},
  url={https://api.semanticscholar.org/CorpusID:258822803}
}

@article{blockfusion,
  title={BlockFusion: Expandable 3D Scene Generation using Latent Tri-plane Extrapolation},
  author={Wu, Zhennan and Li, Yang and Yan, Han and Shang, Taizhang and Sun, Weixuan and Wang, Senbo and Cui, Ruikai and Liu, Weizhe and Sato, Hiroyuki and Li, Hongdong and Ji, Pan},
  journal={ACM Transactions on Graphics},
  volume={43},
  number={4},
  year={2024},
  doi={10.1145/3658188}
 }

@misc{fang2023ctrl,
  title={Ctrl-room: Controllable text-to-3d room meshes generation with layout constraints},
  author={Fang, Chuan and Dong, Yuan and Luo, Kunming and Hu, Xiaotao and Shrestha, Rakesh and Tan, Ping},
  booktitle={2025 International Conference on 3D Vision (3DV)},
  pages={692--701},
  year={2025},
  organization={IEEE}
}

@misc{stan2023ldm3dvr,
      title={LDM3D-VR: Latent Diffusion Model for 3D VR}, 
      author={Gabriela Ben Melech Stan and Diana Wofk and Estelle Aflalo and Shao-Yen Tseng and Zhipeng Cai and Michael Paulitsch and Vasudev Lal},
      year={2023},
      eprint={2311.03226},
      archivePrefix={arXiv},
}

@inproceedings{Structured3D,
  title     = {Structured3D: A Large Photo-realistic Dataset for Structured 3D Modeling},
  author    = {Jia Zheng and Junfei Zhang and Jing Li and Rui Tang and Shenghua Gao and Zihan Zhou},
  booktitle = {Proceedings of The European Conference on Computer Vision (ECCV)},
  year      = {2020},
    doi={10.1007/978-3-030-58545-7_30}
}

@inproceedings{Matterport3D,
  title={Matterport3D: Learning from RGB-D Data in Indoor Environments},
  author={Chang, Angel and Dai, Angela and Funkhouser, Thomas and Halber, Maciej and Nießner, Matthias and Savva, Manolis and Song, Shuran and Zeng, Andy and Zhang, Yinda},
  journal=threedv,
  year={2017},
  pages={667-676},
    doi={10.1109/3DV.2017.00081}
}

@misc{feng2023diffusion360,
  title={Diffusion360: Seamless 360 Degree Panoramic Image Generation Based on Diffusion Models},
  author={Feng, Mengyang and Liu, Jinlin and Cui, Miaomiao and Xie, Xuansong},
  archivePrefix={arXiv},
  eprint={2311.13141},
  year={2023}
}

@article{ozguroglu2024pix2gestalt,
        title={pix2gestalt: Amodal Segmentation by Synthesizing Wholes},
        author={Ege Ozguroglu and Ruoshi Liu and D\'idac Sur\'s and Dian Chen and Achal Dave and Pavel Tokmakov and Carl Vondrick},
        journal=cvpr,
        year={2024},
doi = {10.1109/CVPR52733.2024.00377},
pages = {3931-3940},
    }

@InProceedings{Cohen-Bar_2023_ICCV,
    author    = {Cohen-Bar, Dana and Richardson, Elad and Metzer, Gal and Giryes, Raja and Cohen-Or, Daniel},
    title     = {Set-the-Scene: Global-Local Training for Generating Controllable NeRF Scenes},
    booktitle = {Proceedings of the IEEE/CVF International Conference on Computer Vision (ICCV) Workshops},
    month     = {October},
    year      = {2023},
    pages     = {2920-2929},
    doi={10.1109/ICCVW60793.2023.00314}
}

@InProceedings{Liang_2024_CVPR,
    author    = {Liang, Yixun and Yang, Xin and Lin, Jiantao and Li, Haodong and Xu, Xiaogang and Chen, Yingcong},
    title     = {LucidDreamer: Towards High-Fidelity Text-to-3D Generation via Interval Score Matching},
    booktitle = CVPR,
    month     = {June},
    year      = {2024},
    pages     = {6517-6526},
doi = {10.1109/CVPR52733.2024.00623},
}

@InProceedings{infinite_nature_2020,
  author = {Liu, Andrew and Tucker, Richard and Jampani, Varun and
            Makadia, Ameesh and Snavely, Noah and Kanazawa, Angjoo},
  title = {Infinite Nature: Perpetual View Generation of Natural Scenes from a Single Image},
  booktitle = iccv,
  month = {October},
  year = {2021},
doi = {10.1109/ICCV48922.2021.01419},
}

@INPROCEEDINGS{Factoring2018,
  author={Tulsiani, Shubham and Gupta, Saurabh and Fouhey, David and Efrosefros, Alexei A. and Malik, Jitendra},
  booktitle=cvpr, 
  title={Factoring Shape, Pose, and Layout from the 2D Image of a 3D Scene}, 
  year={2018},
  pages={302-310},
  doi={10.1109/CVPR.2018.00039}}

@inproceedings{armeni20193d,
  title={3D Scene Graph: A Structure for Unified Semantics, 3D Space, and Camera},
  author={Armeni, Iro and He, Zhi-Yang and Gwak, JunYoung and Zamir, Amir R and Fischer, Martin and Malik, Jitendra and Savarese, Silvio},
  booktitle=iccv,
  pages={5663--5672},
  year={2019},
  doi={10.1109/ICCV.2019.00576},
}

@article{po2024state,
      title={State of the Art on Diffusion Models for Visual Computing},
      author={Po, Ryan and Yifan, Wang and Golyanik, Vladislav and Aberman, Kfir and Barron, Jonathan T and Bermano, Amit and Chan, Eric and Dekel, Tali and Holynski, Aleksander and Kanazawa, Angjoo and others},
      journal={Computer Graphics Forum},
      volume={43},
number = {2},
      pages={e15063},
    year={2024},     
doi = {https://doi.org/10.1111/cgf.15063}
    }

@misc{dreamshaper2023,
    title = {DreamShaper v8 Inpainting},
    author = {Lykon},
    year = {2023},
    publisher = {Hugging Face},
    howpublished = {\url{https://huggingface.co/Lykon/dreamshaper-8-inpainting}},
}

@online{polyhaven,
  author       = {PolyHaven},
    title = {PolyHaven},
urldate={2025-01-10},
year={2025},
  url          = {https://polyhaven.com/hdris},
}

@misc{clipscore,
    title = "{CLIPS}core: A Reference-free Evaluation Metric for Image Captioning",
    author = "Hessel, Jack  and
      Holtzman, Ari  and
      Forbes, Maxwell  and
      Le Bras, Ronan  and
      Choi, Yejin",
    editor = "Moens, Marie-Francine  and
      Huang, Xuanjing  and
      Specia, Lucia  and
      Yih, Scott Wen-tau",
    booktitle = "Proceedings of the 2021 Conference on Empirical Methods in Natural Language Processing",
    month = nov,
    year = "2021",
    address = "Online and Punta Cana, Dominican Republic",
    publisher = "Association for Computational Linguistics",
    url = "https://aclanthology.org/2021.emnlp-main.595/",
    doi = "10.18653/v1/2021.emnlp-main.595",
    pages = "7514--7528"
}

@misc{qalign,
  title={Q-Align: Teaching LMMs for Visual Scoring via Discrete Text-Defined Levels},
  author={Wu, Haoning and Zhang, Zicheng and Zhang, Weixia and Chen, Chaofeng and Li, Chunyi and Liao, Liang and Wang, Annan and Zhang, Erli and Sun, Wenxiu and Yan, Qiong and Min, Xiongkuo and Zhai, Guangtai and Lin, Weisi},
  journal={arXiv preprint arXiv:2312.17090},
  year={2023},
  institution={Nanyang Technological University and Shanghai Jiao Tong University and Sensetime Research},
  note={Equal Contribution by Wu, Haoning and Zhang, Zicheng. Project Lead by Wu, Haoning. Corresponding Authors: Zhai, Guangtai and Lin, Weisi.}
}

@inproceedings{clipiqa,
    author = {Wang, Jianyi and Chan, Kelvin CK and Loy, Chen Change},
    title = {Exploring CLIP for Assessing the Look and Feel of Images},
    booktitle = {AAAI},
    year = {2023},
doi = {10.1609/aaai.v37i2.25353},
articleno = {284},
numpages = {9},
}

@misc{yu2024viewcraftertamingvideodiffusion,
      title={ViewCrafter: Taming Video Diffusion Models for High-fidelity Novel View Synthesis}, 
      author={Wangbo Yu and Jinbo Xing and Li Yuan and Wenbo Hu and Xiaoyu Li and Zhipeng Huang and Xiangjun Gao and Tien-Tsin Wong and Ying Shan and Yonghong Tian},
      year={2024},
      eprint={2409.02048},
      archivePrefix={arXiv},
}

@misc{ren2025gen3c3dinformedworldconsistentvideo,
      title={GEN3C: 3D-Informed World-Consistent Video Generation with Precise Camera Control}, 
      author={Xuanchi Ren and Tianchang Shen and Jiahui Huang and Huan Ling and Yifan Lu and Merlin Nimier-David and Thomas Müller and Alexander Keller and Sanja Fidler and Jun Gao},
      year={2025},
      eprint={2503.03751},
      archivePrefix={arXiv},
}

@misc{chen2025flexworldprogressivelyexpanding3d,
      title={FlexWorld: Progressively Expanding 3D Scenes for Flexiable-View Synthesis}, 
      author={Luxi Chen and Zihan Zhou and Min Zhao and Yikai Wang and Ge Zhang and Wenhao Huang and Hao Sun and Ji-Rong Wen and Chongxuan Li},
      year={2025},
      eprint={2503.13265},
      archivePrefix={arXiv}
}

@misc{zhai2025stargenspatiotemporalautoregressionframework,
  title={Stargen: A spatiotemporal autoregression framework with video diffusion model for scalable and controllable scene generation},
  author={Zhai, Shangjin and Ye, Zhichao and Liu, Jialin and Xie, Weijian and Hu, Jiaqi and Peng, Zhen and Xue, Hua and Chen, Danpeng and Wang, Xiaomeng and Yang, Lei and others},
  booktitle={Proceedings of the Computer Vision and Pattern Recognition Conference},
  pages={26822--26833},
  year={2025}
}

@article{SceneSplatter2025,
        title   = {Scene Splatter: Momentum 3D Scene Generation from Single Image with Video Diffusion Model},
        author  = {Zhang, Shengjun and Li, Jinzhao and Fei, Xin and Liu, Hao and Duan, Yueqi},
        journal = {IEEE / CVF Computer Vision and Pattern Recognition Conference (CVPR)},
        year    = {2025},
        url={https://doi.org/10.48550/arXiv.2504.02764}
        }

@misc{kalischek2025cubediffrepurposingdiffusionbasedimage,
      title={CubeDiff: Repurposing Diffusion-Based Image Models for Panorama Generation}, 
      author={Nikolai Kalischek and Michael Oechsle and Fabian Manhardt and Philipp Henzler and Konrad Schindler and Federico Tombari},
      year={2025},
      eprint={2501.17162},
      archivePrefix={arXiv},
      primaryClass={cs.CV},
      url={https://arxiv.org/abs/2501.17162}, 
}

@article{text2lightchen2022,
author = {Chen, Zhaoxi and Wang, Guangcong and Liu, Ziwei},
title = {Text2Light: Zero-Shot Text-Driven HDR Panorama Generation},
year = {2022},
issue_date = {December 2022},
publisher = {Association for Computing Machinery},
address = {New York, NY, USA},
volume = {41},
number = {6},
issn = {0730-0301},
url = {https://doi.org/10.1145/3550454.3555447},
doi = {10.1145/3550454.3555447},
journal = {ACM Trans. Graph.},
month = nov,
articleno = {195},
numpages = {16}
}

@misc{ren2024groundedsamassemblingopenworld,
      title={Grounded SAM: Assembling Open-World Models for Diverse Visual Tasks}, 
      author={Tianhe Ren and Shilong Liu and Ailing Zeng and Jing Lin and Kunchang Li and He Cao and Jiayu Chen and Xinyu Huang and Yukang Chen and Feng Yan and Zhaoyang Zeng and Hao Zhang and Feng Li and Jie Yang and Hongyang Li and Qing Jiang and Lei Zhang},
      year={2024},
      eprint={2401.14159},
      archivePrefix={arXiv},
      url={https://arxiv.org/abs/2401.14159}, 
}

@misc{zhang2023ram,
title={Recognize Anything: A Strong Image Tagging Model},
author={Zhang, Youcai and Huang, Xinyu and Ma, Jinyu and Li, Zhaoyang and Luo, Zhaochuan and Xie, Yanchun and Qin, Yuzhuo and Luo, Tong and Li, Yaqian and Liu, Shilong and others},
archivePrefix={arXiv},
eprint={2306.03514},
year={2023}
}
}

\end{document}